\documentclass[showpacs,preprintnumbers,amsmath,aps,nofootinbib,amssymb,superscriptaddress]{revtex4}
\usepackage{graphicx}
\usepackage{dcolumn}
\usepackage[english]{babel}
\usepackage{bm}
\usepackage[font=scriptsize]{caption}
\usepackage{ragged2e}
\usepackage[font=small,labelfont=bf,justification=raggedright]{caption}
\usepackage[utf8]{inputenc}
\usepackage[tikz]{bclogo}
\usepackage[framemethod=tikz]{mdframed}
\usepackage{amsmath}

\usepackage{mathtools}
\usepackage{xcolor}
\usepackage{lscape}
\usepackage{rotating}
\usepackage{float}
\usepackage{hyperref}
\colorlet{BLUE}{blue} \colorlet{RED}{red}
\newsavebox{\measurebox}

\begin{document}

\title[]{(Pre)-Inflationary Dynamics with Starobinsky Potential in Noncommutative Effective LQC}
\author{L. R. Díaz-Barrón}
\email{lrdiaz@ipn.mx}
\affiliation{Unidad Profesional Interdisciplinaria de Ingenier\'ia, Campus Guana\-jua\-to del Instituto Polit\'ecnico Nacional.\\ Av. Mineral de Valenciana No. 200, Col. Fraccionamiento Industrial Puerto Interior, C.P. 36275, Silao de la Victoria, Guanajuato, M\'exico.}
 
 \author{A. Espinoza-García}
\email{aespinoza@ipn.mx} 
\affiliation{Unidad Profesional Interdisciplinaria de Ingenier\'ia, Campus Guana\-jua\-to del Instituto Polit\'ecnico Nacional.\\ Av. Mineral de Valenciana No. 200, Col. Fraccionamiento Industrial Puerto Interior, C.P. 36275, Silao de la Victoria, Guanajuato, M\'exico.}

\author{S. P\'erez-Pay\'an}
\email{saperezp@ipn.mx} 
\affiliation{Unidad Profesional Interdisciplinaria de Ingenier\'ia, Campus Guana\-jua\-to del Instituto Polit\'ecnico Nacional.\\ Av. Mineral de Valenciana No. 200, Col. Fraccionamiento Industrial Puerto Interior, C.P. 36275, Silao de la Victoria, Guanajuato, M\'exico.}

\author{J. Socorro}
\email{socorro@fisica.ugto.mx} 
\affiliation{Departamento de Física, DCeI, Universidad de Guanajuato-Campus León, León, 37150, Guanajuato, México}

\begin{abstract}
In this work, we investigate the (pre)-inflationary dynamics of a flat, homogeneous, and isotropic universe governed by the Starobinsky potential within the framework of noncommutative effective loop quantum cosmology. The field equations are solved numerically for various initial conditions and different values of the noncommutative parameter. We analyze the background dynamics for three representative regimes—the extreme kinetic-energy domination, kinetic-energy domination, and potential-energy domination. A complementary analysis is performed from the viewpoint of dynamical systems, highlighting the qualitative features of the scalar field evolution. Finally, a discussion comparing our results with previous studies employing the chaotic (quadratic) potential in the same formalism is presented.
\end{abstract}

\maketitle

\section{Introduction} 
The standard cosmological model, based on Einstein's theory of General Relativity (GR) and the hot Big Bang paradigm, successfully accounts for many observed features of the universe. However, it leaves several fundamental issues unresolved, such as the flatness and horizon problems, and the origin of the large-scale structure. The inflationary scenario --a period of accelerated expansion in the very early universe-- offers a compelling mechanism to address these puzzles \cite{Starobinsky:1980te, Guth:1980zm, Linde:1981mu}. 

Although inflation can be consistently implemented within GR, the classical theory is expected to break down near the Planck scale, where quantum-gravitational effects become significant. A consistent quantum-gravity framework is therefore required to understand the earliest stages of cosmic evolution and to set the initial conditions for inflation. loop quantum gravity (LQG) \cite{Rovelli:2004tv, Thiemann:2007pyv, Ashtekar:2004eh} provides a promising approach, predicting a discrete geometry at the Planck scale. Loop quantum cosmology (LQC) applies the quantization techniques of LQG to homogeneous cosmological settings, yielding a non-singular evolution in which the Big Bang singularity is replaced by a quantum bounce that connects a contracting pre-Big Bang phase to an expanding post-bounce universe \cite{Ashtekar:2006rx, Ashtekar:2006wn, Bojowald:2008zzb}. This framework motivates a new paradigm in which a non-singular pre-inflationary era precedes inflation.

In this context, several works have examined the inflationary dynamics in LQC, including models that incorporate the Starobinsky potential — one of the most observationally favored potentials according to Planck data \cite{Planck:2015sxf}. For example, Bonga and Gupt \cite{Bonga:2015kaa} explored Starobinsky inflation in standard LQC, while Bhardwaj {\it et al.}\cite{Bhardwaj:2018omt} provided a detailed study of slow-roll inflation and its consistency with loop quantum corrections. Further avenues within this line of inquiry have been explored in \cite{Bonga:2015xna,Li:2020mfi,Li:2019qzr,Saeed:2024xhk}.

At the beginning of the 20th century, interest was renewed in the concept of non-commutative spacetime \cite{Connes:1997cr, Seiberg:1999vs}, initially proposed by Snyder \cite{Snyder:1946qz, Snyder:1947nq} as a means to regularize quantum field theory. Later developments in string theory and LQG suggested that spacetime may possess a fundamentally discrete structure, for which noncommutative geometry offers an effective description. Although most studies focus on fixed, flat backgrounds, there have also been attempts to incorporate noncommutativity directly into gravitational models \cite{Moffat:2000gr, Chamseddine:2000zu, Garcia-Compean:2003nix,Garcia-Compean:2002dgp, Kober:2014wsa}. Because full noncommutative gravity is mathematically intricate, simplified minisuperspace approaches have been developed \cite{Garcia-Compean:2001jxk, Barbosa:2004kp}, inspiring numerous extensions to cosmological and astrophysical contexts \cite{Guzman:2007zza, Aguero:2007pg, Guzman:2008gz, Perez-Payan}.

Building upon these developments, the present work continues to investigate noncommutative effective loop quantum cosmology, first introduced in \cite{Espinoza-Garcia:2017qjl}. Subsequent studies \cite{Diaz-Barron:2019awc, Diaz-Barron:2021yha,Diaz-Barron:2023ctp} established the noncommutative deformation of effective LQC equations and demonstrated that the fundamental features of LQC, such as the bounce, are preserved. At the same time, novel phenomena can appear, including emergent inflationary potentials. Recent analyses \cite{Diaz-Barron:2021yha,Diaz-Barron:2023ctp} considered a quadratic potential (chaotic), finding that when the noncommutative parameter is small, the dynamics closely reproduce those of standard LQC. Other related studies have examined pre-inflationary behavior within noncommutative modified LQC \cite{Mohammadi:2024lgo}.

Investigating the dynamics of the pre-inflationary epoch is crucial for establishing the theoretical framework necessary to properly analyze primordial cosmological perturbations and their associated noncommutative modifications. Extensions to the standard scenario for cosmological perturbations have been extensively developed, particularly within the context of (LQC). Specifically, in references \cite{Dressed1, Dressed2}, the classical background geometry is supplanted by a fully quantum description, yielding an effective, non-singular ``dressed" geometry whose effects are imprinted on the perturbations. Furthermore, modifications incorporating canonical spacetime noncommutativity have been proposed, as detailed in \cite{Hybrid,Akofor:2007fv}. Such rigorous theoretical analyses are expected to provide significant insight into persistent observational challenges, notably the $H_0$ tension \cite{reviewsH0,reviewsH0_1,reviewsH0_2,reviewsH0_3,reviewsH0_4,reviewsH0_5,reviewsH0_6,reviewsH0_7}. Clarifying this discrepancy—which lies between local measurements (such as those from Supernovae and Cepheids) and inferences drawn from the Cosmic Microwave Background (CMB) within the standard $\Lambda$CDM model—is a key objective. The equation of state parameter is recognized as playing a potentially critical role in this resolution. However, the complete implementation and detailed analysis of this proposed framework is beyond the scope of the present article and is reserved for subsequent future work.

In this paper, we employ the noncommutative effective loop quantum cosmology (NC-ELQC) framework to study pre-inflationary dynamics of a flat, homogeneous, and isotropic universe driven by a scalar field with a Starobinsky potential. The manuscript is organized as follows. Section \ref{ncelqc} provides a concise review of the noncommutative effective LQC formalism. Section \ref{nc_elqc_Starobinsky} derives the field equations for the Starobinsky potential and presents numerical solutions for the different energy regimes, namely the extreme kinetic energy domination (EKED), the kinetic energy domination (KED) and the potential energy domination (PED). Section \ref{dynamical_system} presents a complementary dynamical-systems analysis. In Section \ref{discusion} a comparative analysis between the Starobinsky potential and the quadratic potential, within the NC-ELQC formalism, is presented. Finally, Section \ref{conclusions} summarizes our conclusions.

\section{Brief Review of Noncommutative Effective LQC}\label{ncelqc}
\subsection{Effective Loop Quantum Cosmology}

We commence by delineating several key aspects of the effective scheme of loop quantum cosmology, for which comprehensive and detailed discussions are available in the standard reviews  \cite{Bojowald:2008zzb,Ashtekar:2011ni}. As previously indicated, LQC constitutes the quantization of symmetry-reduced general relativity, adhering closely to the foundational principles and methodologies of loop quantum gravity. In brief, the conventional 3-metric Hamiltonian formulation of General Relativity is reformulated, via a canonical transformation, in terms of the SU(2) Ashtekar-Barbero connection, thereby recasting General Relativity as a gauge theory \cite{Bojowald:2008zzb,Ashtekar:2011ni}. Furthermore, the rigorous implementation of canonical quantization necessitates the adoption of holonomies of the Ashtekar-Barbero connection as the appropriate configuration variables, in place of the connection itself. Consequently, holonomies pertaining to a highly symmetric sector (e.g., the homogeneous sector) of the connection superspace are treated as the cosmologically relevant configuration variables. Canonical quantization is subsequently carried out on this connection minisuperspace, resulting in the quantum framework known as loop quantum cosmology. Nevertheless, LQC inherits the same limitations associated with the minisuperspace approximation as standard quantum cosmology; specifically, LQC does not represent a symmetry-reduced version of LQG. Notably, the noncommutativity among the conjugate momentum fields (electric fluxes) present in full LQG is entirely absent in LQC, indicating a potentially significant loss of information due to the minisuperspace reduction.

A principal achievement of LQC is the resolution of the initial cosmological singularity through the mechanism of a quantum bounce. Quantum corrections within this framework are primarily categorized as holonomy- and inverse-triad-related, with holonomy corrections generally playing a more significant role. The standard effective scheme of LQC is constructed by emphasizing these holonomy-induced quantum corrections. In the isotropic setting, the classical geometro-dynamical variables designated by the effective LQC approach are $\beta=\gamma\dot{a}/(aN)$ and $V=a^3$, which satisfy:
\begin{equation}
\{\beta,V\}=4\pi G\gamma,\quad
\mathsf{P}(y)=4\pi G\gamma\, \partial_{\beta}\wedge\partial_{V}\label{PB},
\end{equation}
where $\gamma$ is the so-called Barbero–Immirzi parameter \cite{Ashtekar:2004eh}, and $\mathsf{P}(y)$ represents the Poisson structure in the variables $y = (\beta, V)$. The gravitational Hamiltonian, expressed in these variables for $N = 1$, takes the form
\begin{equation}
\mathcal{H}=-\frac{3}{8\pi G\gamma^{2}}\beta^2V.
\label{ham1}
\end{equation}
It has been demonstrated, through both analytical \cite{Bojowald:2008zzb,Ashtekar:2011ni} and numerical~\cite{Diener:2014hba,Diener:2017lde} studies, that in the isotropic case, the quantum corrections of LQC are accurately captured by implementing the 
\textit{replacement} 
\begin{equation}
\beta\to\frac{\sin{\lambda\beta}}{\lambda},
\end{equation}
within the classical Hamiltonian, where $\lambda^{2}=4\sqrt{3}\pi\gamma\ell^{2}_{p}$ denotes the lowest eigenvalue of the area operator in full loop quantum gravity (specifically, for quantum states compatible with the assumed spatially homogeneous and isotropic geometry), and $\ell_p$ represents the Planck length. The resulting effective Hamiltonian is expressed as
\begin{equation}
\mathcal{H}_{\mathrm{eff}}=-\frac{3}{8\pi G\gamma^{2}\lambda^{2}}\sin^{2}(\lambda\beta)V+\frac{p^{2}_{\phi}}{2V}.
\label{ham2}
\end{equation}
Thus, the effective framework of LQC facilitates the incorporation of substantive quantum corrections within an otherwise classical setting. This approach has permitted the systematic exploration of quantum effects across a broad spectrum of cosmological backgrounds, all within a tractable yet idealized theoretical context. Recall also that in arriving at the classical Hamiltonian (2) a space integration has been performed on a finite region of space (which is permitted due to the symmetries of the FLRW spacetime), whose volume $V_0$ has been here set to unity, without loss of generality, as is usually done. (Of course, one could work with a generic $V_0$, see the Review \cite{Ashtekar:2011ni} for more details.) Therefore, the physical volume of the universe is given by $V=a^3V_0$. Additionally, it can be established  that the function $\beta$ has dimensions of inverse length (while $\lambda^2$ has dimensions of area), so that the effective Hamiltonian (4) is well defined and, in particular, has dimensions of energy (see \cite{Ashtekar:2006wn, Ashtekar:2011ni}).

The equations of motion corresponding to the “holonomized” Hamiltonian, as given in Eq.~\eqref{ham2}, endowed with the standard Poisson structure,
\begin{equation}
\mathsf{P}(y)=4\pi G\gamma\, \partial_{\beta}\wedge\partial_{V}+\partial_{\phi}\wedge\partial_{p_\phi},\quad y=(\beta,V,\phi,p_\phi),
\label{PB2}
\end{equation}
are
\begin{align}
&\dot{\beta}=4\pi G\gamma\frac{\partial\mathcal{H}_{\mathrm{eff}}}{\partial V}=-\frac{3}{\gamma\lambda^{2}}\sin^{2}(\lambda\beta),\label{eom-beta}\\
&\dot{V}=-4\pi G\gamma\frac{\partial\mathcal{H}_{\mathrm{eff}}}{\partial\beta}=\frac{3}{\gamma\lambda}V\sin(\lambda\beta)\cos(\lambda\beta),\label{eom-vol}\\
&\dot{\phi}=\frac{\partial\mathcal{H}_{\mathrm{eff}}}{\partial p_{\phi}}=\frac{p_{\phi}}{V},\label{eom-phi}\\
&\dot{p}_{\phi}=-\frac{\partial\mathcal{H}_{\mathrm{eff}}}{\partial\phi}=0.\label{eom-p-phi}
\end{align}
Since $V=a^{3}$, $\frac{\dot{V}}{3V}=\frac{\dot{a}}{a}=:H$, then, taking into account the effective Hamiltonian constraint, $\mathcal{H}_{\mathrm{eff}}=0$, and the equation of motion for $V$, we have,
\begin{equation}
H^{2} :=\left(\frac{\dot{V}}{3V}\right)^{2}=\frac{8\pi G}{3}\rho\left(1-\frac{\rho}{\rho_{\mathrm{c}}}\right),
\label{mod-friedmann}
\end{equation}
where 
\begin{equation}
\rho=\frac{\dot\phi^2}{2}=\frac{p^{2}_{\phi}}{2V^{2}}=\frac{3}{8\pi G\gamma^{2}\lambda^{2}}\sin^{2}(\lambda\beta),
\end{equation} 
and $\rho_{{c}}$ is the maximum value that $\rho$ can take in view of the effective Hamiltonian constraint, 
\begin{equation}
\rho_{{c}}=\frac{3}{8\pi G\gamma^{2}\lambda^{2}}\simeq0.41\rho_{\mathrm p},
\end{equation}
with $\rho_{\mathrm p}$ the Planck density. Equation \eqref{mod-friedmann} is the so called \textit{modified Friedmann equation}.

In the context of full loop quantum cosmology, it has been demonstrated that the parameter $\beta$ assumes values within the interval $\left(0, \frac{\pi}{\lambda}\right)$ \cite{Ashtekar:2004eh}. Consequently, the volume function attains a stationary point at $\beta = \frac{\pi}{2\lambda}$. Analysis of the equation of motion~(\ref{eom-beta}) establishes that $\beta$ is a monotonically decreasing function of time ($\dot{\beta} \leq 0$), which in turn implies that this stationary point is encountered only once during the evolution (and, correspondingly, that the maximum of the energy density function is also realized only once). This monotonic behavior of $\beta$ further indicates that the stationary point corresponds to a minimum, as $\rho$ is always bounded, and the stationary point associated with a vanishing volume does not arise in the solutions of the equations of motion. Thus, within the effective framework of LQC, a minimum of the volume function is reached precisely once, and this minimum volume coincides with the maximum of the energy density function. This scenario is succinctly encapsulated by the assertion that singularity resolution is effected via a (minimum-volume) quantum bounce. As previously noted, this singularity resolution constitutes one of the defining features of LQC. Within this effective scheme, it is a generic property: all solutions exhibit a minimum-volume bounce exactly once. Although one might contend that singularity resolution is a consequence of the semiclassical assumptions inherent to this effective framework, it has been demonstrated that the quantum bounce remains "robust" in full LQC, specifically for the flat FLRW model with a free scalar field~\cite{Ashtekar:2007em}.

It is noteworthy that the modified Friedmann equation~(\ref{mod-friedmann}) encapsulates the occurrence of the cosmological bounce—in particular, at $\rho = \rho_{\mathrm{c}}$—in a relatively straightforward manner. In the limiting case $\lambda \to 0$ (i.e., vanishing area gap), the standard Friedmann equation, ${H}^{2} = \frac{8\pi G}{3}\rho$, is recovered.

By employing the Hamiltonian constraint, the equation governing $\beta$ may be decoupled and subsequently solved via direct integration. Substituting the resultant solution $\beta = \beta(t)$ into Eq.~(\ref{eom-vol}) yields the corresponding solution for $V = V(t)$. Analogously, inserting the solution $V = V(t)$ into Eq.~(\ref{eom-phi}) provides the solution $\phi = \phi(t)$. Explicitly,
\begin{align}
&\beta(t)=\frac{1}{\lambda}\mathrm{arccot}\left(\frac{3t}{\gamma\lambda}\right),\label{sol-beta-lqc}\\
&V(t)=C_1\sqrt{\gamma^{2}\lambda^{2}+9t^{2}},\label{sol-v-lqc}\\
&\phi(t)=C_2+\frac{p_{\phi}}{3C_1}\mathrm{log}\left(3t+\sqrt{\gamma^2\lambda^2+9t^2}\right),\label{sol-fi-lqc}\\
&p_\phi(t)=p_\phi.
\end{align}
From (\ref{sol-beta-lqc}) we note that the solution is consistent with $\beta\in\left(0,\frac{\pi}{\lambda}\right)$ (by considering another ``branch'' we get $\beta\in\left[\left.-\frac{\pi}{2\lambda},\frac{\pi}{2\lambda}\right.\right)$, which is also employed in the literature of effective LQC (see, for instance, \cite{corichi-tatiana}).

\subsection{Noncommutative Effective LQC}
Deformed minisuperspace models \cite{Garcia-Compean:2001jxk}, were initially developed to address the complexities of a complete noncommutative theory of gravity. By incorporating a deformation directly into the minisuperspace, these models effectively simulate noncommutative effects. In canonical quantum cosmology, the Wheeler-DeWitt equation—derived via canonical quantization—typically exhibits a Klein-Gordon-like structure, serving as the framework for the quantum dynamics of the universe.

An alternative and powerful method for investigating quantum effects involves deforming the system's phase space. This technique, known as deformation quantization \cite{Cordero:2011xa}, offers an equivalent perspective by viewing quantum mechanics as a noncommutative generalization of classical mechanics. Focusing on cosmological models, which are typically studied in the minisuperspace approximation, a modified phase space allows for deeper insight into quantum effects in cosmological solutions \cite{Vakili:2010qf}.

This modification of the phase space is achieved by introducing deformations via the Moyal brackets $\{f,g\}_{\alpha}=f\star_{\alpha} g-g\star_{\alpha} f$, where the standard product of functions is replaced by the Moyal product \cite{Moyal} $(f\star g)(x)=\exp[\frac{1}{2}\alpha^{ij}\partial_{i}^{(1)}\partial_{j}^{(2)}]f(x_1)g(x_2)\vert_{x_1=x_2=x}$. The deformation parameter $\alpha_{ij}$ is represented by the matrix \cite{Djemai:2003kd}
\begin{eqnarray}
\alpha_{ij} =
\left( {\begin{array}{cc}
 \theta_{ij} & \delta_{ij}+\sigma_{ij}  \\
- \delta_{ij}-\sigma_{ij} & \eta_{ij}  \\
 \end{array} } \right).
\end{eqnarray}
The resulting noncommutative phase space is characterized by the $2\times 2$ antisymmetric matrices $\theta_{ij}$ and $\eta_{ij}$, representing the noncommutativity between the coordinates and momenta, respectively. The parameter $\sigma_{ij}$ is defined by the expression $\sigma_{ij}=-1/8(\theta_{i}^k\eta_{kj}+\eta_{i}^k\theta_{kj})$. The corresponding deformed phase space algebra is given by
\begin{equation}\label{def_algebra}
\{x_i,x_j\}_{\alpha}=\theta_{ij}, \;\{x_i,p_j\}_{\alpha}=\delta_{ij}+\sigma_{ij},\; \{p_i,p_j\}_{\alpha}=\eta_{ij}.
\end{equation}
In the rest of this paper we will use the particular deformations $\theta_{ij}=-\theta\epsilon_{ij}$ and $\eta_{ij}=\eta\epsilon_{ij}$, where $\epsilon_{ij}$ is the Levi-Civita symbol. Restricting ourselves to a 4-dimensional phase space, an alternative method for achieving an algebraic structure similar to (\ref{def_algebra}) is through a linear canonical transformation of the phase space variables. This transformation can be expressed as
\begin{eqnarray}\nonumber
\hat{x}=x+\frac{\theta}{2}p_{y}, \qquad \hat{y}=y-\frac{\theta}{2}p_{x},\\
\hat{p}_{x}=p_{x}-\frac{\eta}{2}y, \qquad \hat{p}_{y}=p_{y}+\frac{\eta}{2}x, \label{nctrans1}
\end{eqnarray}
where $\{x,y,p_x,p_y\}$ obey the standard Poisson algebra, leading to a deformed algebra 
\begin{equation}
\{\hat{y},\hat{x}\}=\theta,\; \{\hat{x},\hat{p}_{x}\}=\{\hat{y},\hat{p}_{y}\}=1+\sigma,\; \{\hat{p}_y,\hat{p}_x\}=\eta,\label{dpa}
\end{equation}
with $\sigma=\theta\eta/4$. It is noteworthy that Eq.~(\ref{nctrans1}) employs standard Poisson brackets, while  Eq.~(\ref{def_algebra}) utilizes $\alpha$-deformed Poisson brackets.

Two primary approaches are viable for formulating the deformed theory. Firstly, the $\alpha$-deformed algebra (\ref{def_algebra}) can be applied directly to a system with a canonical Hamiltonian $\mathcal H$, which induces a deformed (non-canonical) structure in the equations of motion via the $\alpha$-deformed Poisson brackets. Alternatively, one can use the "shifted variables" introduced by the transformation in (\ref{nctrans1}). In this second approach, the equations of motion retain their canonical structure, while the Hamiltonian $\mathcal{H}_{nc}$—possessing the same functional form as $\mathcal{H}$—is expressed in terms of these shifted variables, which satisfy the algebra (\ref{dpa}). In this work, we adopt the latter approach, consistent with previous studies in various cosmological scenarios \cite{Vakili:2010qf, Perez-Payan, Perez-Payan:2014kea, Lopez:2017xaz, Malekolkalami:2014dca}.

We now introduce a modified algebraic structure, analogous to Eq.~(\ref{dpa}), within the phase space of effective LQC, defined by the variables  ($\beta, \phi, V, p_{\phi}$). Specifically, we consider the following modified algebra in the momentum sector
\begin{equation}
\{\beta^{nc},V^{nc}\}=4\pi G\gamma,\quad \{V^{nc},p^{nc}_{\phi}\}=\theta, \quad\{\phi^{nc},p^{nc}_{\phi}\}=1,
\label{ncm-free}
\end{equation}
where all other Poisson brackets are zero. This modified algebra can be realized by employing the shifted variables
\begin{equation}\label{ps_trans}
\beta^{nc}=\beta, \quad \phi^{nc}=\phi, \quad V^{nc}=V+a'\phi\theta, \quad p_{\phi}^{nc}=p_\phi+b\beta\theta,
\end{equation}
while $a'$ and $b$ are dimensionless constants arising from the generalization of the linear canonical transformations of the phase-space variables, Eq. (\ref{nctrans1}), subject to the constraint $a'-4\pi G\gamma b=1$. Since $a'$ and $b$ are dimensionless constants, the non-commutative parameter $\theta$ possesses  units of $l_{pl}^3/m_{pl}$.

The construction of the deformed theory begins with the formulation of an effective Hamiltonian. This Hamiltonian is formally analogous to the canonical Hamiltonian (\ref{ham2}), but is constructed using the shifted variables subject to the algebraic relations in (\ref{ncm-free}). Consequently, the noncommutative effective Hamiltonian is
\begin{equation}\label{ham_eff_nc}
\mathcal{H}_{eff}^{nc}=-\frac{3V^{nc}}{8\pi G\gamma^2\lambda^2}\sin^2(\lambda\beta)+\frac{(p_{\phi}^2)^{nc}}{2V^{nc}}+\mathcal{V}(\phi)V^{nc}.
\end{equation}
This Hamiltonian explicitly includes a generic potential term $\mathcal{V}(\phi)$. Reference \cite{Diaz-Barron:2021yha} has shown that by taking the subset of the transformation (\ref{ps_trans}) characterized by $b=0$, $V^{nc}(t_c)>0$, the system defined by (\ref{ham_eff_nc}) achieves a single minimum volume bounce at $t=t_c$ where the energy density is $\rho(t_c)=\rho_c$. Furthermore, a prior analysis of the pre-inflationary dynamics for the quadratic potential was conducted in \cite{Diaz-Barron:2023ctp}, accounting for these restrictions. 

\section{Starobinsky Inflation in Noncommutative Effective LQC}\label{nc_elqc_Starobinsky}
This section presents a detailed analysis of the pre-inflationary and post-bounce dynamics using the effective Hamiltonian (\ref{ham_eff_nc}) when the Starobinsky potential governs the scalar field. Our investigation focuses on the constraints  $b=0$, $V^{nc}(t_c)>0$, which are known to ensure a single minimum-volume bounce at a time $t=t_c$, where the energy density reaches the critical value $\rho(t_c)=\rho_c$.
\subsection{Field Equations and Modified Friedmann Equation}
The matter source is a scalar field whose self-interaction potential drives the inflationary phase \cite{Barrow:1988xi, Starobinsky:2001xq, DeFelice:2010aj}
\begin{equation}\label{Starobinsky_pot}
\mathcal{V}(\phi)=\frac{3m^2}{32\pi G}\left( 1-e^{-\sqrt{\frac{16\pi G}{3}}\phi} \right)^2,
\end{equation}
this is the well-known Starobinsky potential , where $m=2.49\times 10^{-6}~m_{pl}$ is the mass of the field. 

The equations of motion derived from the noncommutative effective Hamiltonian (\ref{ham_eff_nc}) are
\begin{eqnarray}
\label{feg1} \dot{\beta}=4\pi G\gamma\left[-\frac{3}{4\pi G\gamma^2\lambda^2}\sin^2(\lambda\beta)+2 \mathcal{V}(\phi) \right],\label{field_eqn_1}\\
\label{feg2} \dot{V}=\frac{3}{\gamma\lambda}V^{{nc}}\sin(\lambda\beta)\cos(\lambda\beta)-\frac{4\pi G\gamma b\theta p_{\phi}^{{nc}}}{V^{{nc}}},\label{field_eqn_2}\\
\label{feg3} \dot{\phi}=\frac{p_{\phi}^{{nc}}}{V^{{nc}}},\label{field_eqn_3}\\
\label{feg4} \dot{p}_{\phi}=\frac{3a\theta}{4\pi G\gamma^{2}\lambda^{2}}\sin^{2}(\lambda\beta)-2a\theta \mathcal{V}(\phi)-V^{{nc}}\partial_\phi \mathcal{V}(\phi)\label{field_eqn_4}.
\end{eqnarray}
In the limit where the noncommutative parameter $\theta\to 0$, the standard commutative equations are recovered. Analysis of the equations of motion of the scalar field $\phi$ and the effective Hamiltonian, $\mathcal{H}_{eff}^{nc}$, reveals that the energy density is $\rho=\frac{\dot\phi^2}{2}+\mathcal{V}(\phi)=\rho_c\sin^2(\lambda\beta)$, where this functional dependence on $\beta$ is identical to that observed in the standard LQC scenario, thereby ensuring that the same maximum energy density, $\rho_c$, is attained. Nevertheless, the explicit functional dependence on the cosmic time, $t$, will generally differ from the standard case, as the solution for $\beta(t)$ is anticipated to be contingent upon the noncommutative parameter $\theta$. Despite this, given the established monotonic behavior of $\beta$, the overarching qualitative behavior of the energy density function is expected to parallel that of the standard case. It is further noted that, in principle, the volume function could attain multiple stationary values, potentially giving rise to several bouncing scenarios. The realization of these scenarios depends on the occurrence and evolution of all dynamical variables.

The equation of motion governing the volume, $V$, suggests that the simultaneous satisfaction of the conditions 
\begin{equation}
b=0,\qquad V^{nc}(t_c)>0,\label{condition_b=0}
\end{equation}
where $t_c$ is defined by the energy density reaching its maximum value, $\rho(t_c)=\rho_c$, constitutes a sufficient condition for a minimum volume bounce to be reached precisely when the energy density $\rho$ achieves its maximum value $\rho_c$. This condition, while already implicit in the free case, was not explicitly articulated in the previous literature \cite{Espinoza-Garcia:2017qjl}. Specifically, condition (\ref{condition_b=0}) preserves the standard LQC big bounce. Furthermore, this scenario represents the sole bouncing event, given that $\beta$ is a monotonically decreasing function of time (as indicated by equation (\ref{feg1})) with values confined to the interval $(0,\frac{\pi}{\lambda})$, and because, $V^{nc}=0$ is strictly prohibited during the evolution, consequence of the scalar field equation of motion and the bounded nature of the energy density.

Conversely, the simultaneous fulfillment of the conditions 
\begin{equation}
\dot\phi(t_c)=0,\qquad b\theta\ddot{\phi}(t_c)<0, \label{condition_b2}
\end{equation}
which is equivalent to
\begin{equation}
\rho(\phi(t_c))=\mathcal{V}(\phi(t_c)),\qquad b\theta\ddot{\phi}(t_c)<0, \label{condition_b3}
\end{equation}
also constitutes a sufficient condition for the occurrence of minimum-volume bounce. In this case, multiple stationary points of $V(t)$ may appear, and the singularity resolution is not always guaranteed in this latter scenario. Condition (\ref{condition_b2}) would serve to specify the value of the scalar field at its minimum volume bounce. Moreover, the scalar field's equation of motion mandates that $p_{\phi}(t_c)=-b\theta\beta(t_c)=-\frac{b\theta\pi}{2\lambda}$ must be satisfied. This leads to the following: by requiring a minimum-volume bounce at the maximum energy density, but without imposing the particular noncommutativity representation where $b=0$, the initial conditions for all dynamical variables (except the volume) are fixed at the bounce. This stipulation results in a drastically reduced of solutions that exhibit a single bounce event somewhat analogous to the single big bounce of standard LQC. This outcome is evidently less desirable from a general theoretical perspective.

It's imperative to note that while conditions (\ref{condition_b=0}) and (\ref{condition_b2}) do not exhaust the entirety of possibilities for the existence of a minimum volume bounce, they do represent a complete set of possibilities for a minimum volume bounce occurring specifically at the maximum energy density $\rho=\rho_c$. Furthermore, these conditions are mutually exclusive; no solution that features such a bouncing event can satisfy them simultaneously.

The generalized Klein-Gordon and continuity equations can describe the dynamics of the scalar field, formulated using Equations (\ref{feg1})-(\ref{feg4}). These expressions incorporate contributions from both loop quantum cosmology and noncommutative geometry
\begin{eqnarray}
\label{nc_KG} \ddot\phi+3H^{nc}\dot\phi+\partial_{\phi}\mathcal{V}(\phi)=0,\\
\label{nc_cont} \dot\rho+3H^{nc}(\rho+P)=0,
\end{eqnarray}
here, the effective Hubble parameter is defined as $H^{nc}\equiv\frac{\dot{V}^{nc}}{3V^{nc}}$, and the energy density $\rho$ and pressure $P$ are $\rho=\frac{\dot\phi^2}{2}+\mathcal{V}(\phi)$ and $P=\frac{\dot\phi^2}{2}-\mathcal{V}(\phi)$, respectively. As expected, from the limit $\theta\to 0$ we recover the commutative counterparts of Equations (\ref{nc_KG}) and (\ref{nc_cont}).

Finally, a modified Friedmann equation that incorporates corrections from both LQC and noncommutativity can be derived \cite{Diaz-Barron:2019awc}
\begin{equation}\label{Fnc}
H^2=\frac{8\pi G}{3}\rho\left[1-\frac{\rho+\rho_\theta}{\rho_c}\right]. 
\end{equation}
In this equation $\rho_\theta=\rho_1(\theta)+\rho_2(\theta^2)$ represents the noncommutative corrections terms given by
\begin{equation}
\rho_1(\theta)=\frac{2a\theta\phi}{V}(\rho_c-\rho)+\frac{\sqrt 2b\theta}{\gamma\lambda V}\sqrt{\frac{\rho_{c}}{\rho}\left(1-\frac{\rho}{\rho_c}\right)(\rho-\mathcal{V}(\phi))},
\end{equation}
\begin{equation}
\begin{split}
\rho_2(\theta^2)=\frac{a^2\theta^2\phi^2}{V^2}(\rho_c-\rho)&+\frac{\sqrt 2ab\theta^2\phi}{\gamma\lambda V}\sqrt{\frac{\rho_{c}}{\rho}\left(1-\frac{\rho}{\rho_c}\right)(\rho-\mathcal{V}(\phi))}\\
&+\frac{4\pi Gb^2\theta^2}{3V^2}\rho_c\left(1-\frac{\mathcal{V}(\phi)}{\rho}\right).
\end{split}
\end{equation}
Due to the complex form of the Friedmann equation (\ref{Fnc}), a numerical approach is imperative for the subsequent analysis. In the specific scenario where the parameter $b=0$, the noncommutative correction to the energy density, $\rho_\theta$, is directly proportional to the difference term $(\rho-\rho_c)$. Consequently, at the critical energy density $\rho=\rho_c$, this noncommutative contribution identically vanishes, and the bounce condition remains unaltered.

\subsection{Energy Stages}
The inflationary evolution is classified into three energy regimes by defining the ratio of the potential energy at the bounce $\mathcal V(\phi_B)$ to critical energy density $\rho_c$
\begin{equation}
F_B=\frac{\mathcal V(\phi_B)}{\rho_c},
\end{equation}
where the value of $F_B$ ranges from $[0,1]$, and depends on the initial conditions, determining the energy stage immediately following the bounce, as discussed in \cite{Ashtekar:2009mm, Ashtekar:2011rm}. Specifically, if $F_B<10^{-4}$, the system is in the extreme kinetic energy domination regime. For $10^{-4}<F_B<0.5$, the system is in the kinetic energy domination regime. Finally, when $0.5<F_B<1$, the system is in the potential energy domination regime.

A physical quantity of interest is the effective equation of state (EoS) parameter for the scalar field, $\omega_{\phi}$, defined as the ratio of pressure to energy density 
\begin{equation}\label{eos}
\omega_{\phi}=\frac{P(t)}{\rho(t)}=\frac{\frac{1}{2}\dot\phi^2-\mathcal{V}(\phi)}{\frac{1}{2}\dot\phi^2+\mathcal{V}(\phi)},
\end{equation}
 and describes how the pressure of a component of the universe relates to its energy density. This value determines how that component affects the expansion of the universe. The parameter, $\omega_{\phi}$, can take values between $[-1,1]$, with $\omega_{\phi}=-1$ corresponding to the PED regime, and $\omega_{\phi}=1$ corresponding to the KED regime. Consistent with the stages identified in standard LQC \cite{Li:2019ipm, Zhu:2017jew}, the scalar field evolution before reheating can be classified into three distinct phases: the bouncing stage, the transition stage, and the slow-roll inflationary stage. The bouncing phase includes a period of super-inflation (SI), characterized by an increasing Hubble parameter. This epoch spans from the initial expansion, the bounce, until the Hubble parameter reaches its maximum value, hence $\dot H=0$.

To ensure both accelerated expansion and a gradual rate of inflation, the slow-roll parameters must satisfy the conditions  $\epsilon_{H}\ll 1$ and $\eta_{H}\ll 1$. These parameters are defined as \cite{prfl}
\begin{equation}
\epsilon_{H}=-\frac{\dot H}{H^2}, \qquad \eta_{H}=-\frac{\ddot H}{2\dot{H} H},
\end{equation}
and the number of e-folds corresponding to this period is 
\begin{equation}
N=\ln\left(\frac{a_{end}}{a_i}\right),
\end{equation}
where $a_{end}$ is the value of $a$ at the end of slow-roll inflation and $a_i$ is the value at the onset. The requirement for an early expansion consistent with observations demands that $N\geq 60$ \cite{Planck:2018jri}.

Initial investigations of inflation in noncommutative effective LQC \cite{Diaz-Barron:2021yha} established that for a generic potential, an inflationary solution ensuring a bounce requires $b=0$ and $p_\phi=\frac{b\theta\pi}{2\lambda}$. However, the analysis was specialized to the quadratic potential, leading to the condition $b=0$ for a satisfactory inflationary period. For the Starobinsky potential studied in this work, both initial conditions are considered relevant.

\subsection{EKE Domination: After the Bounce}

\begin{figure}[H]
  \includegraphics[width=.47\linewidth]{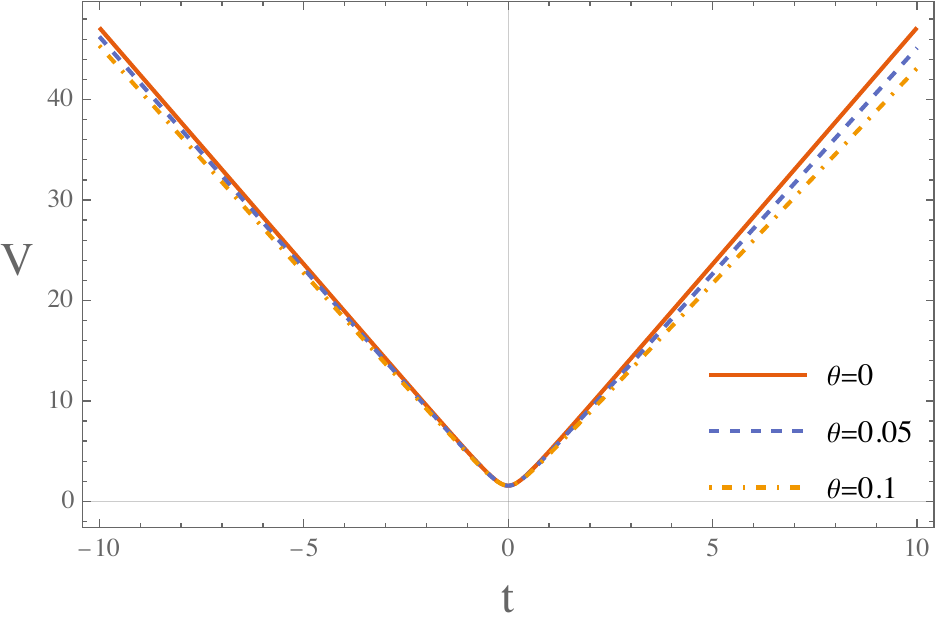} \hfill
  \includegraphics[width=.47\linewidth]{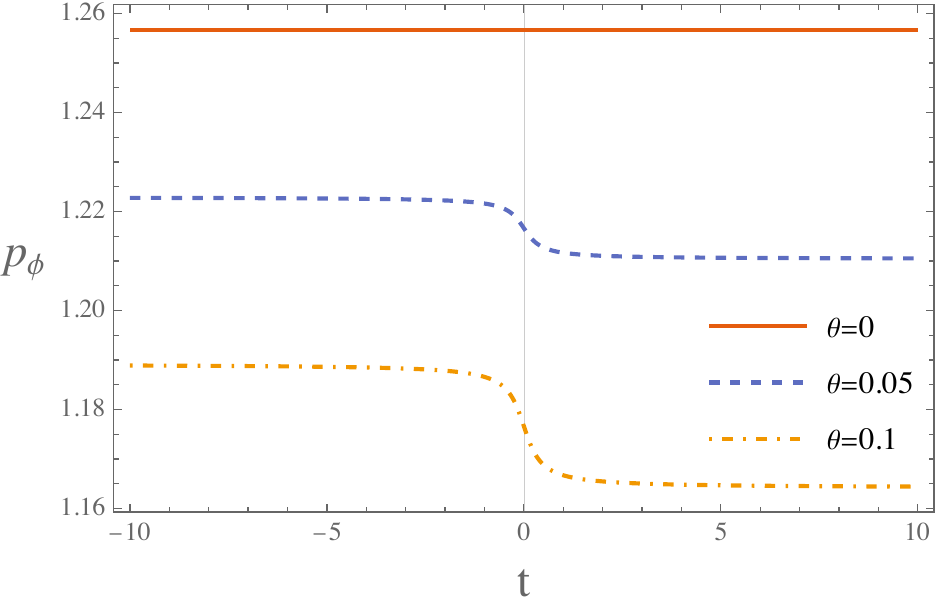}
\caption{The figure shows the numerical solutions of the volume function (left panel) and the conjugate momentum of the scalar field (right panel), in the EKED stage. It is observed that, for both $V$ and $p_{\phi}$, an increase in $\theta$ diminishes their values relative to their commutative analogs.}
\label{fig:figs_solEKED}
\end{figure}

\begin{figure}[H]
  \includegraphics[width=.47\linewidth]{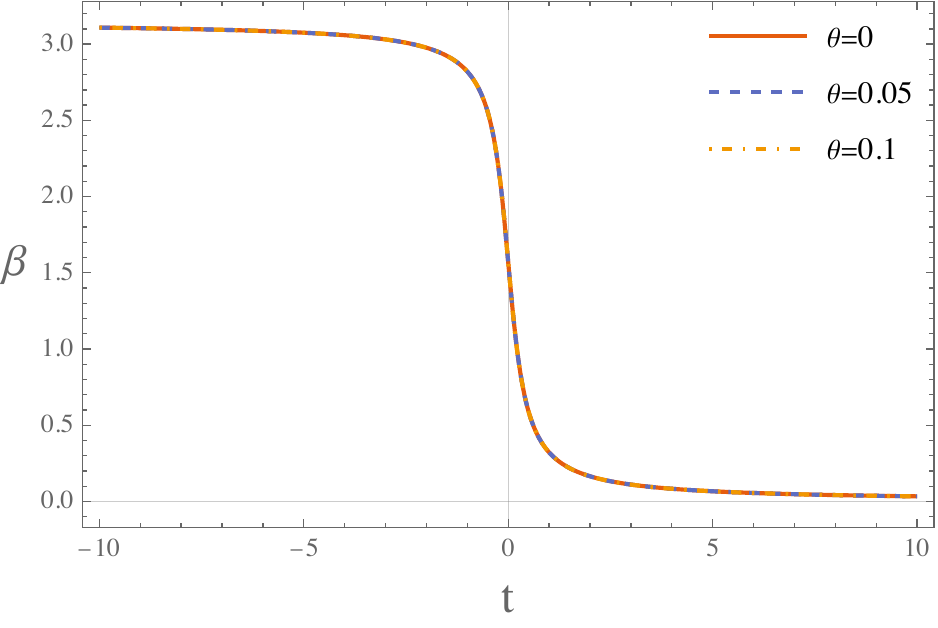}\hfill
  \includegraphics[width=.47\linewidth]{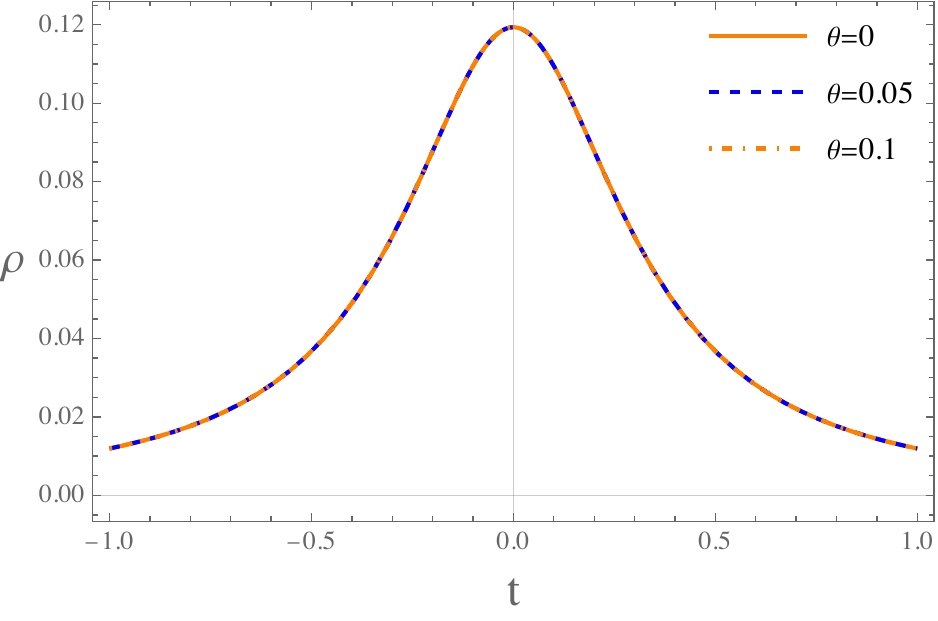}
  \caption{Behavior of $\beta$ (left panel) and the energy density $\rho$ (right panel), in the EKED stage. The numerical solutions show that $\beta$ is independent of the noncommutative parameter, and consequently $\rho$ exhibits a behavior identical to the commutative case.}
\label{fig:figs_beta_rho_EKED}
\end{figure}

\begin{figure}[H]
  \centering
  \includegraphics[width=.7\linewidth]{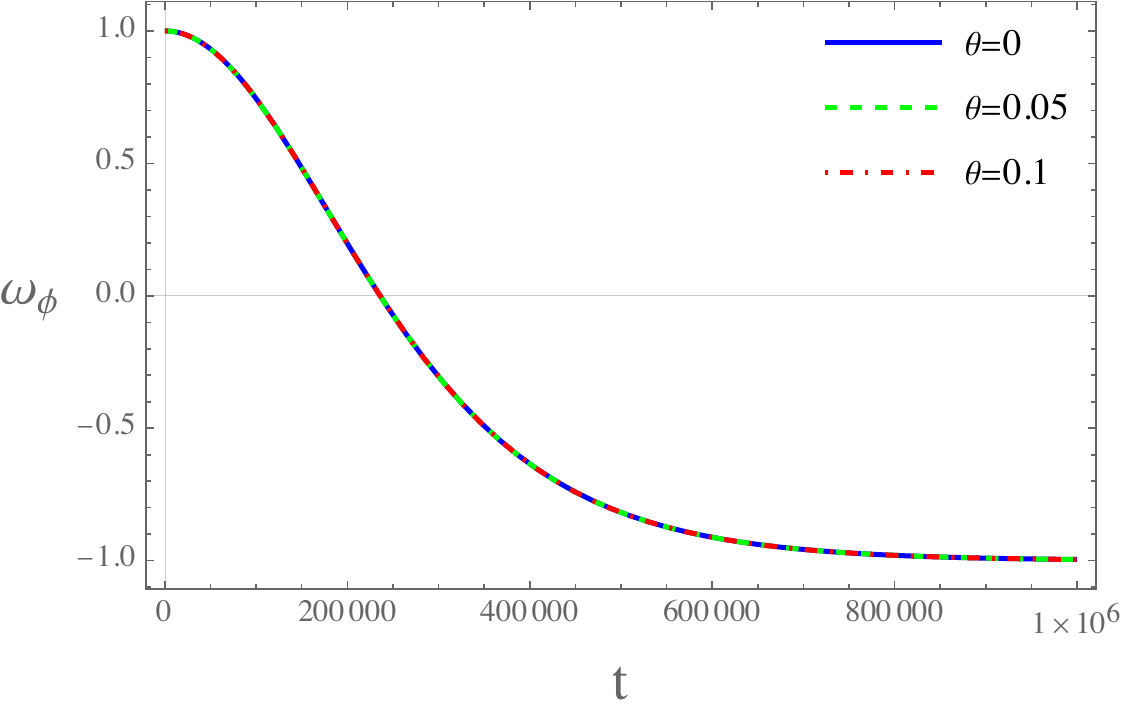 }
  \caption{The figure illustrates the behavior of the effective equation of state parameter, $\omega_{\phi}$. Within the regime of extreme kinetic energy domination, we can identify distinct stages: the bounce stage, the transition stage, and the inflationary stage.}
  \label{fig:omega_EKED}
\end{figure}

We begin with the initial conditions for the extreme kinetic energy domination after the bounce: $ F_B\le10^ {-4}$, $\phi_B=1~m_{pl}$, and $\dot\phi_B=0.8~m_{pl}^2$. For our numerical simulations, we used the numerical integration scheme in Mathematica 14. The numerical solutions corresponding to the field equations (\ref{field_eqn_1}-\ref{field_eqn_4}) are presented in Fig~\ref{fig:figs_solEKED}, where $\gamma,\lambda, G=1$ and $\rho_c=\frac{3}{8\pi}$, will be retained throughout the subsequent numerical computations. The left panel of figure \ref{fig:figs_solEKED} clearly illustrates the noncommutative corrections to the volume, where an increase in the noncommutative parameter results in a decrease of the volume relative to its commutative counterpart. This same behavior is observed in the solution for the $p_{\phi}$ field, as depicted in the right panel of Fig.~\ref{fig:figs_solEKED}.
The solution for $\beta$, as shown in the left-hand side of Fig.~\ref{fig:figs_beta_rho_EKED}, is independent of the noncommutative parameter. Consequently, the energy density regardless of the value of $\theta$, behaves identically to the commutative case, which can be observed in the right-hand side of figure \ref{fig:figs_beta_rho_EKED}.
From the evolution of the effective EoS parameter, shown in figure \ref{fig:omega_EKED}, several stages can be identified: the bounce period, the transition period, and the inflationary stage, in accordance with \cite{Li:2019ipm}, which corresponds to the (pre)-inflationary dynamics of the universe.
\subsubsection{Bouncing Phase}\label{bouncing}
\begin{figure}[H]
  \includegraphics[width=.47\linewidth]{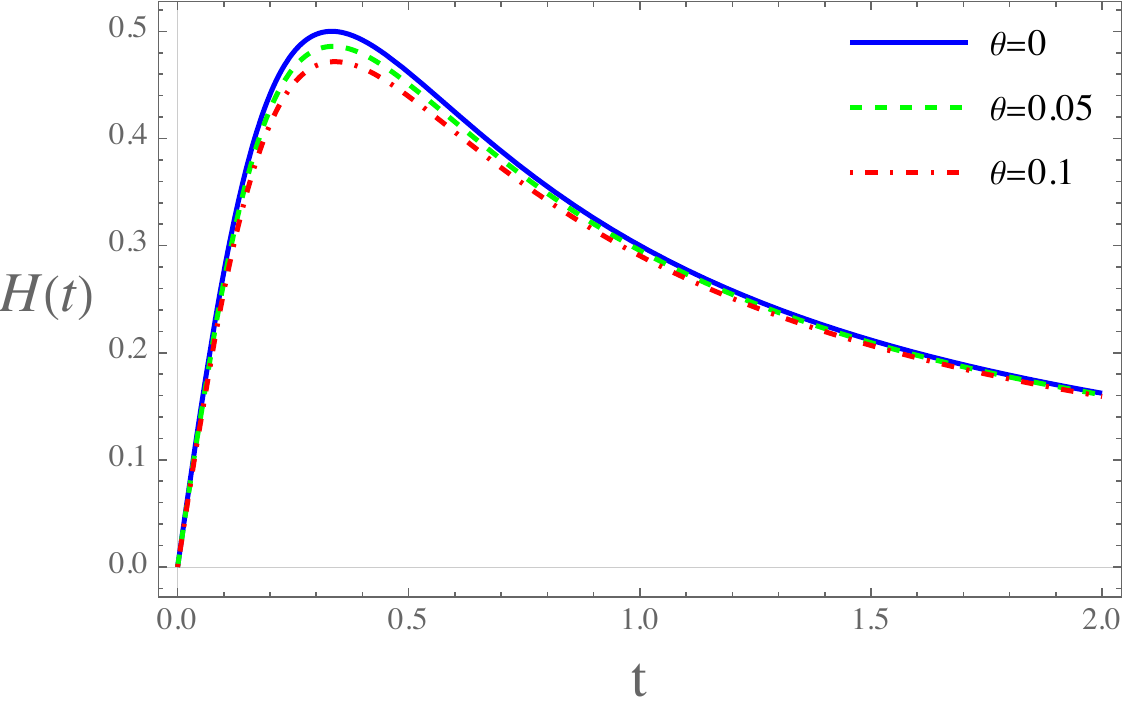}\hfill
   \includegraphics[width=.47\linewidth]{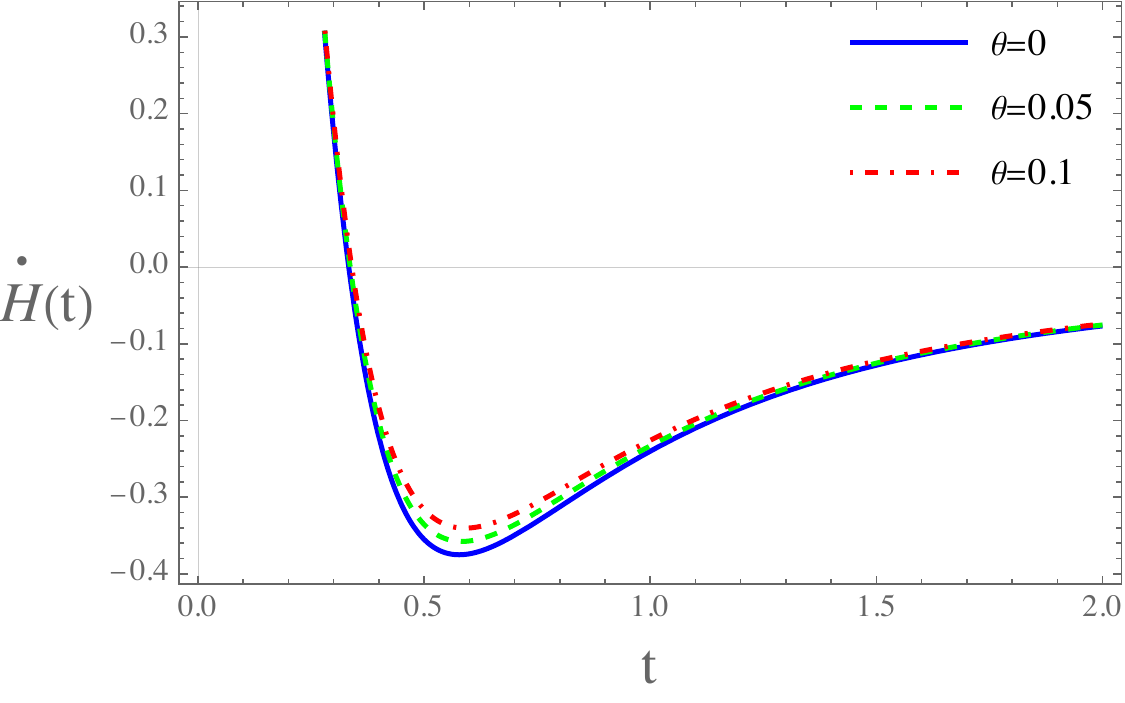}
 \caption{The figure illustrates the behavior of the Hubble parameter, as well as its corresponding rate of change. In the left panel, we can observe that noncommutativity reduces the duration of the SI phase. The right panel shows that the SI phase concludes when $\dot H=0$.}
\label{fig:figs_EKED1}
\end{figure}

The bouncing stage begins at the bounce and ends when $\omega_{\phi}=0$, this occurs, regarding the effective EoS parameter (\ref{eos}), when the scalar field's kinetic energy equals its potential energy. We will label these two moments as $t_B^{EKED}$ and $t_{EQ}^{EKED}$ respectively. In Table~\ref{tab:EKED_Bouncing}, we present the data obtained from numerical solutions for the commutative and noncommutative models. For the commutative case, $\theta=0$, the bouncing stage comes to an end at $t_{EQ}^{EKED}=2.359\times10^{5}$. This phase also includes a period of super-inflation (SI), during which the Hubble parameter experiences extremely rapid growth, reaching a maximum $H_{EQ}^{EKED}=0.5$. Also, at the end of this phase, the effects of LQG corrections can be neglected, recovering the classical Friedmann dynamics. In the left panel of figure \ref{fig:figs_EKED1}, we can observe the behavior of the Hubble parameter in the SI stage. From the right panel of Fig.~\ref{fig:figs_EKED1}, we can see that the SI period comes to an end when $\dot{H}=0$, that is, when $t_{SI}^{EKED}=0.3333$, and it lasts $N_{SI}^{EKED}=0.115525$ e-folds.
\begin{table*}[t]
			\scriptsize
				\caption{Numerical results of the noncommutative model of LQC with the Starobinsky potential in the EKED stage. The Events: bounce, the end of SI (super-inflation), the equilibrium point KE=PE, the onset of slow-roll inflation, and the end of inflation, are compared for $\theta=0$, $\theta=0.05$, and $\theta=0.1$.}
				\label{tab:EKED_Bouncing}
\begin{center}
				\begin{tabular}{c   l  c  c  c  c  c c}
					\hline
					$\bf{b=0}$ & \bf Event & $\boldsymbol t$ & $\boldsymbol{\phi}$ & $\boldsymbol{\dot\phi}$ &$\bf H$ & $ \bf\dot H$ & $\bf N$ \\ \hline
					\hline
					&Bounce & 0 & 1 & 0.8  & $6.123\times10^{-17}$& 3& 0\\
					&End SI &0.3333 & 0.235033 & 0.565686 & 0.5 &$2.4121\times10^{-16}$ & 0.115525  \\
					$\theta=0$& KE=PE & $2.359\times 10^{5}$(\footnote{~the numerical results show a difference of $10^{-1}$ with respect to noncommutative cases}) & 3.76032 &$9.9599\times10^{-7}$ & $1.7607\times10^{-6}$ & $-4.6500\times10^{-12}$& 4.52763 \\
					& Onset Slow Roll & $1.2692\times10^7$ & 3.99535 &$-3.2025\times10^{-14}$ & $1.245\times10^{-6}$ &$2.33281\times10^{-24}$& 20.1027 \\  
					& End of Inflation & $ ---$ & $---$ & $ ---$ & $ ---$ & $ ---$ & $---$\\ \hline
					& Bounce & 0 & 1 & 0.8  & $6.123\times10^{-17}$& 3 &0\\
					& End SI & 0.332787 & 0.235245 & 0.568538 & 0.497346 & $-1.86424\times10^{-16}$ &  0.114793  \\
					$\theta=0.05$\ & KE=PE & $2.359\times10^{5}{~(^a)}$ & 3.78972 & $1.00449\times10^{-6}$ & $1.7607\times10^{-6}$ & $-4.6500\times10^{-12}$ & 4.5230379\\
					& Onset Slow Roll & $4.41877\times 10^6$ & 4.02676 &$1.08404\times10^{-13}$ & $1.245\times10^{-6}$ &$1.03675\times10^{-25}$& 9.79795 \\ 
					& End of Inflation & $ ---$ & $---$ & $ ---$ & $ ---$ & $ ---$ & $---$\\ \hline
					& Bounce & 0 & 1 & 0.8  & $6.123\times10^{-17}$& 3&0\\
					& End SI & 0.332211 & 0.235446 & 0.571462 & 0.494674 & $4.53962\times10^{-17}$& 0.114048  \\
					$\theta=0.1$\ & KE=PE & $2.359\times10^5{~(^a)}$ & 3.82029 & $1.01334\times10^{-6}$ & $1.76069\times10^{-6}$ &$-4.650\times10^{-12}$& 4.5230377 \\
					& Onset Slow Roll & $3.58045\times10^6$ & 4.05942 &$1.90968\times10^{-13}$ & $1.245\times10^{-6}$ &$5.22759\times10^{-22}$& 9.64405 \\ 
					& End of Inflation & $ ---$ & $---$ & $ ---$ & $ ---$ & $ ---$ & $---$\\ \hline
				\end{tabular}
			\end{center}
		\end{table*}
		
The noncommutative effects at this stage depend on the evolution of the scalar field. Within this period of super-inflation, depicted in the right panel of Fig.~\ref{fig:figs_EKED1}, noncommutativity effects are observed to reduce the duration of this phase. For $\theta=0.05$, this reduction correlates with a contracted growth in the Hubble parameter, where the maxima is attained when $H_{SI_{0.05}}^{EKED}=0.497346$ and the termination of the SI phase occurs when $t_{SI_{0.05}}^{EKED}=0.332787$. When $\theta=0.1$, the super-inflation period concludes at $t_{SI_{0.1}}^{EKED}=0.332211$, with an even lower maximum of the Hubble parameter of $H_{SI_{0.1}}^{EKED}=0.494674$. 

In contrast, during the SI period, we observed that the bouncing stage concludes almost identically in both commutative and noncommutative scenarios, that is, at $t_{EQ}^{EKED}=2.359\times 10^5$ (with differences of order $10^{-1}$). These data are presented in Table~\ref{tab:EKED_Bouncing}.
\subsubsection{Transition Phase}\label{trasition}
This stage is characterized by a sharp decline in kinetic energy, causing the universe to accelerate, while the effective equation of state tends to remain constant. This phase initiates when the kinetic energy is equal to the potential energy $\omega_{\phi}=0$, and terminates when the equation of state parameter acquires a value $\omega_{\phi}=-1$, as shown in figure (\ref{fig:omega_EKED}). For both the commutative and noncommutative cases, this stage starts at $t_{EQ}^{EKED}=2.359\times 10^5$. However, for the commutative case ends at $t_i^{EKED}=1.2692\times10^7$ ($t_i^{EKED}$ will also be the time where the slow roll inflation begins), and for the noncommutative cases $\theta=0.05$ and $\theta=0.1$ ends at $t_{i_{0.05}}^{EKED}=4.41877\times 10^6$ and at $t_{i_{0.1}}^{EKED}=3.58045\times10^6$, respectively.

Similar to the bouncing phase, the noncommutative effects reduced the duration of the transition stage. For instance, for $\theta=0$ the evolution of the e-folds, from the bounce to the end of the transition phase, are $N_{tr}=4.52763$, while for $\theta=0.05$ the number of e-folds gives $N_{tr_{0.05}}^{EKED}=4.5230379$ and for $\theta=0.1$ we have $N_{tr_{0.1}}^{EKED}=4.5230377$ (see Table~\ref{tab:EKED_Bouncing}).
\subsubsection{Slow-Roll Inflation}\label{slowroll}
\begin{figure}[H]
  \includegraphics[width=.47\linewidth]{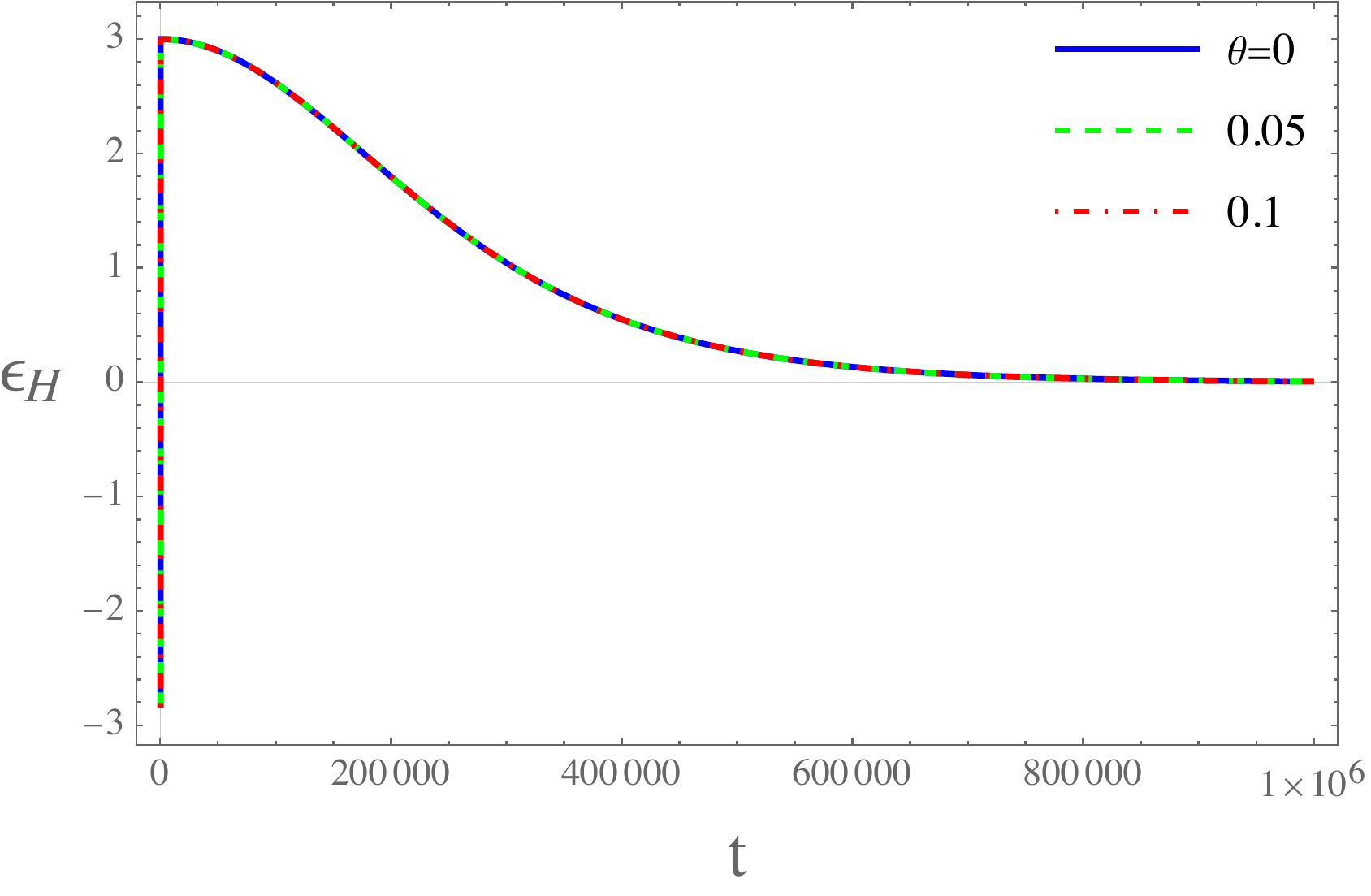}\hfill
   \includegraphics[width=.47\linewidth]{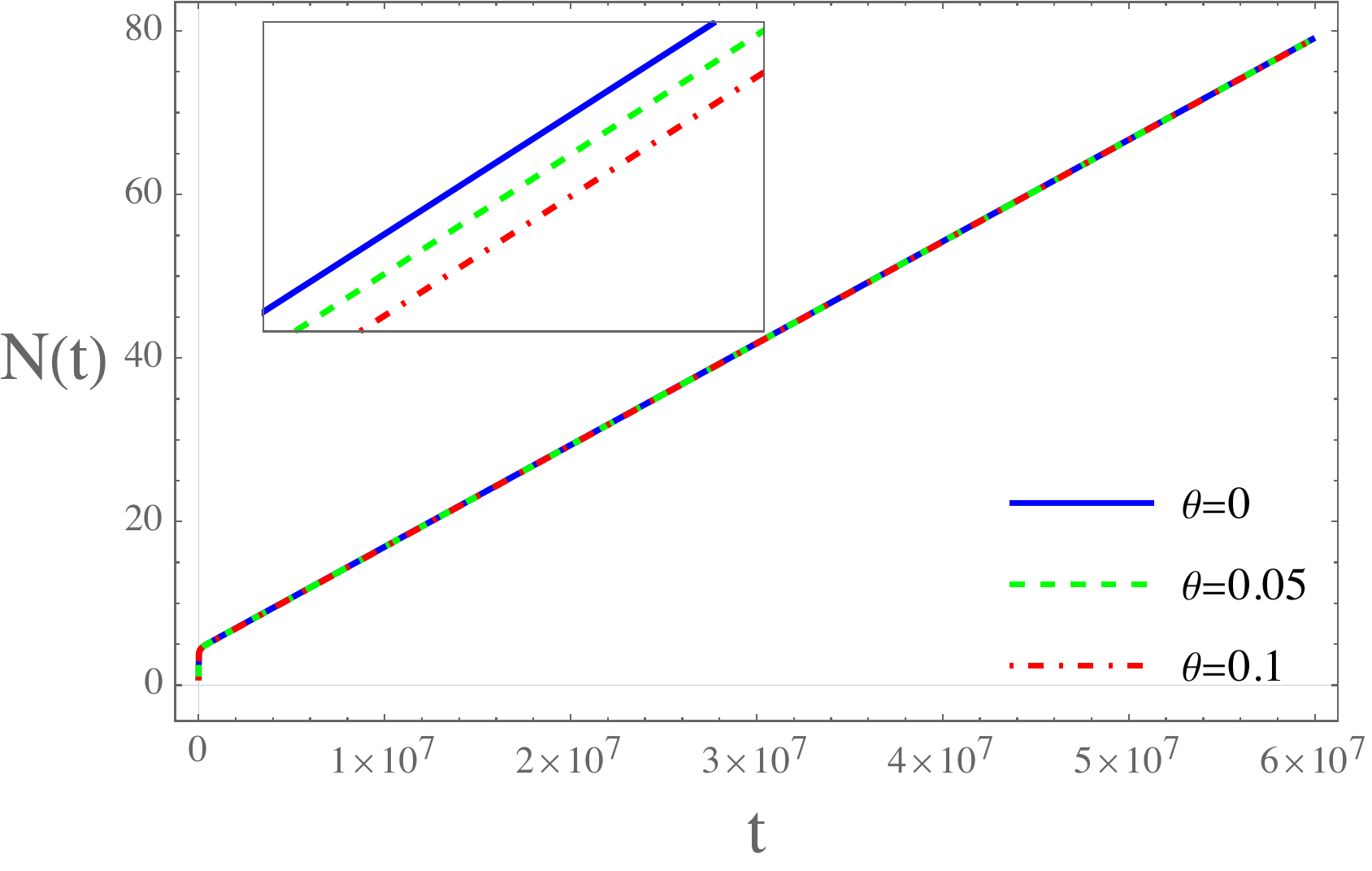}
 \caption{In the figure we show the behavior of the slow-roll parameter $\epsilon_H$ (left panel) and the evolution of the e-folds (right panel).}
\label{fig:figs_eta_efolds}
\end{figure}

\begin{figure}[H]
\centering
  \includegraphics[width=.8\linewidth]{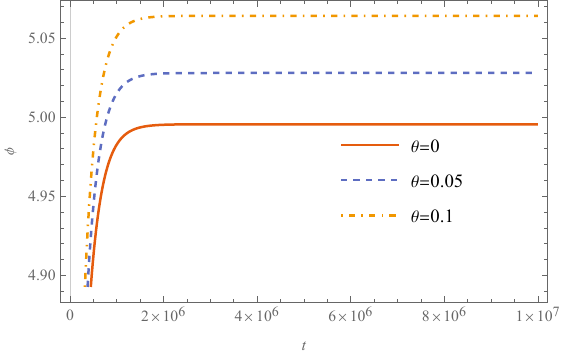}
 \caption{In the figure show the numerical solution of $\phi(t)$ with the initial conditions $\phi_B=1~m_{pl}$ and $\dot\phi_B=0.8~m_{pl}^2$.}
\label{fig:New_phiEKED}
\end{figure}

The onset of the slow-roll inflation stage is marked when $\omega_{\phi}=-1$. At this initial point, the system's energy density is overwhelmingly dominated by the Starobinsky potential (\ref{Starobinsky_pot}). For the commutative case, this transition happens at $t_i^{EKED}=1.2692\times10^7$ with a corresponding field value of $\phi(t_i)=3.99535~m_{pl}$. The left panel of Fig.~\ref{fig:figs_eta_efolds} shows the inflationary region, complying with the condition $\epsilon\ll 1$. Contrary to the behavior of the $\phi^2$ potential, for the initial conditions $(\phi_B=1~m_{pl}$, $\dot\phi_B=0.8~m_{pl}^2)$ the end of this stage never occurs, allowing the inflationary period to be eternal. This observation is evidenced by the initial conditions of the scalar field, which is identified with the inflaton. Commencing at an initial value of $\phi_B=1~m_{pl}$ and possessing an initial positive velocity, $\dot\phi_B>0$, the inflaton field is significantly distant from the minimum of the potential. As depicted in Fig.~\ref{fig:New_phiEKED}, the field value increases until attains a constant value. This behavior implies that the field never descends toward the minimum of the potential, thereby resulting in the phenomenon of eternal inflation.
Although we do not have a timeframe for which the slow-roll inflationary stage comes to an end, from right panel of figure \ref{fig:figs_eta_efolds} we can observed that if we take $t_{sr}^{EKED}=4.5\times 10^7$ the corresponding e-folds for the commutative case are $N_{sr}^{EKED}=60.3261$, while for the noncommutative cases gives $N_{sr_{0.05}}^{EKED}=60.3106$ and $N_{sr_{0.1}}^{EKED}=60.2944$, for $\theta=0.05$ and $\theta=0.1$, respectively. These results are in agreement with the observational data \cite{Planck:2018jri}, which set a bound of $N\geq 60$ for an early inflationary expansion. Throughout the evolution of e-folds, the commutative and the noncommutative cases have a very similar behavior, however, as shown in the enlarged section of the right side of Fig.~(\ref{fig:figs_eta_efolds}), clearly shows that $N(t)$ decreases as the noncommutative parameter takes on larger values. The latter implies that we must have that $\theta\ll 1$ for the universe to evolve to the present time.
\subsection{KE Domination: After the Bounce}
\begin{figure}[H]
  \includegraphics[width=.47\linewidth]{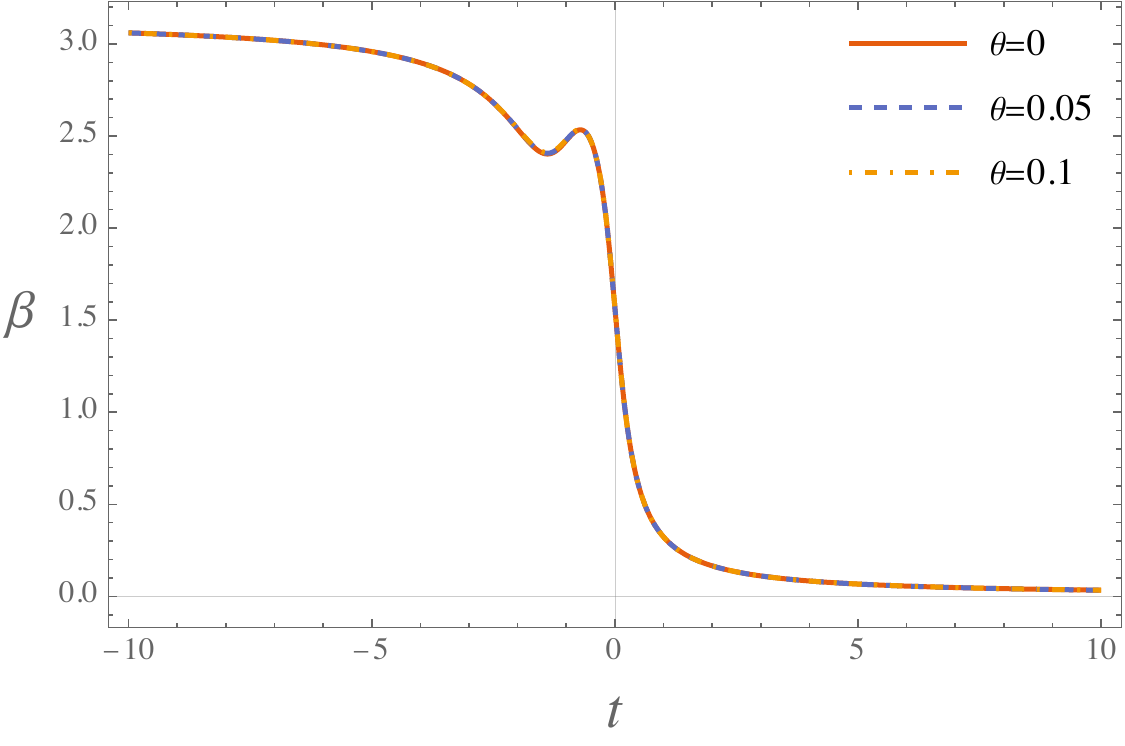}\hfill
   \includegraphics[width=.47\linewidth]{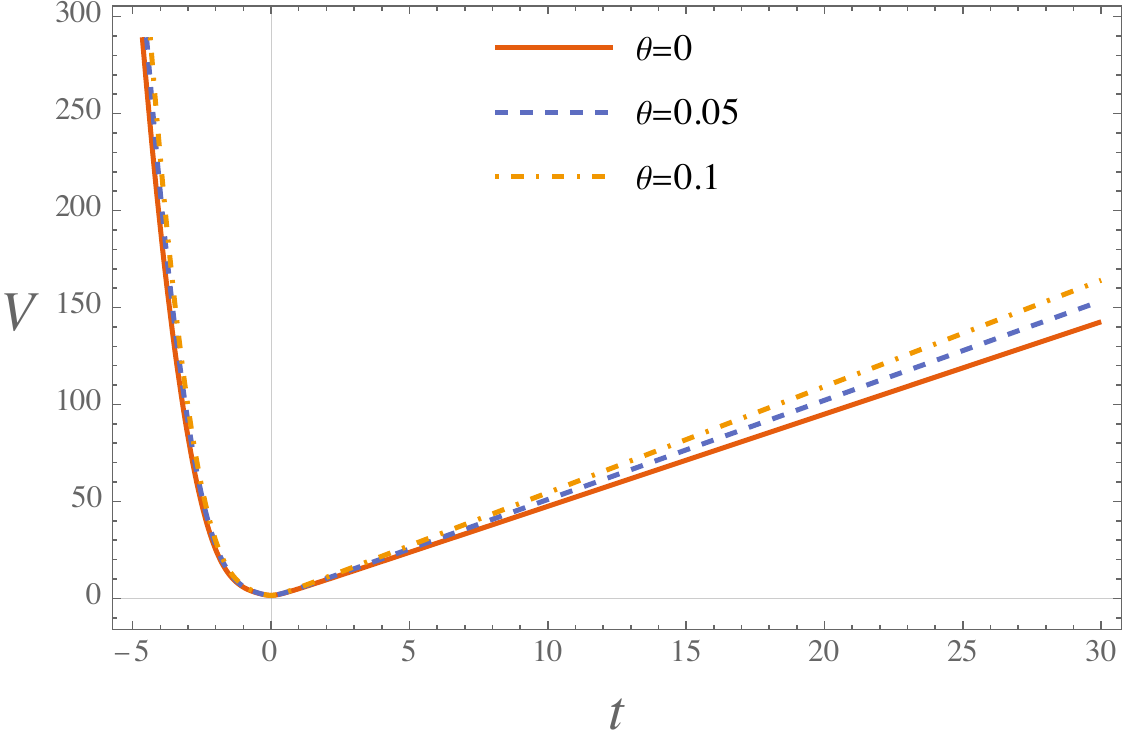}
   \centering
   \includegraphics[width=.47\linewidth]{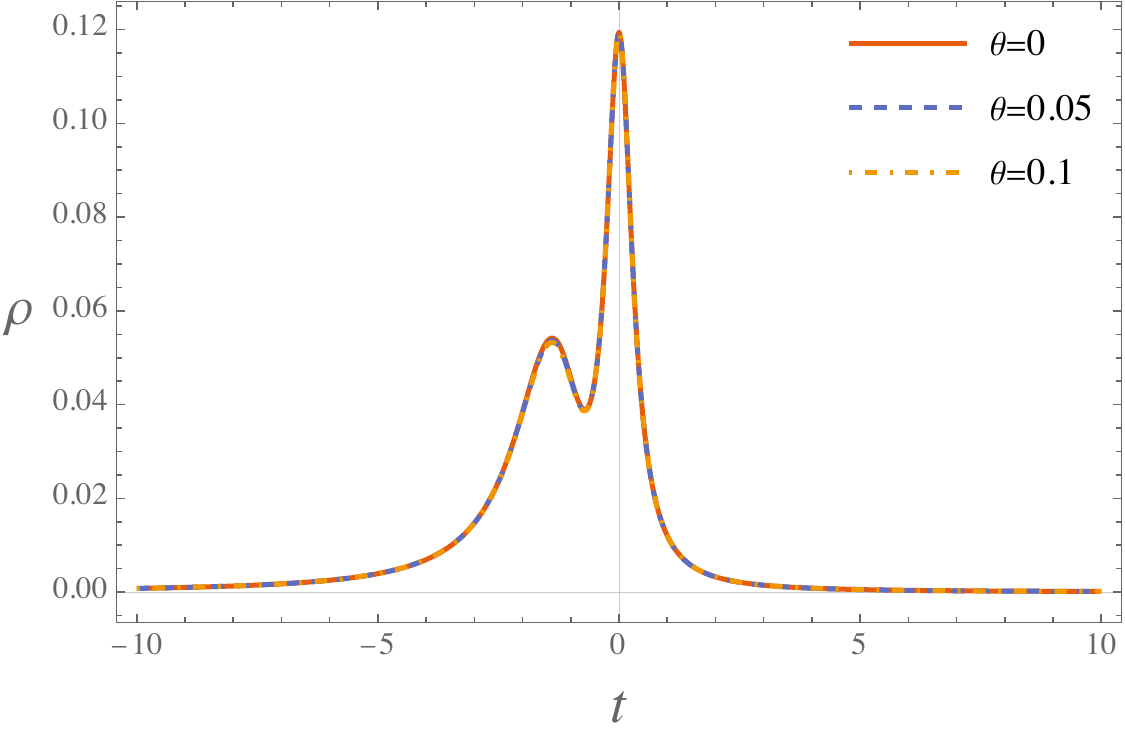}
 \caption{The figure illustrates the behavior of $\beta$, the volume and the energy density. From the top left panel, we can see that $\beta$ is affected by a fluctuation before the bounce. The top right panel shows that the volume function has an asymmetric behavior (contrary to the EKED case). Finally, in the bottom panel, we observed that $\rho$ has a spike which dissipates to then reach its maximum value.}
\label{fig:figs_KED}
\end{figure}

For this stage, the ratio of the potential energy to the critical density is restricted to the values $10^{-4}<F_B<0.5$. To ensure that the system exists within a kinetic energy dominated phase, the initial conditions of the scalar field $\phi_B$ field are required to be negative. Consequently we will have that $\phi_B=-2.8~m_{pl}$ and $\dot\phi=0.8~m_{pl}^2$. Figure \ref{fig:figs_KED} displays the numerical solutions for the field equations for $\beta$ and $V$, as well as the behavior for the energy density $\rho$. In top left panel of Fig.~\ref{fig:figs_KED}, it is clear that the solution for $\beta$ is affected by a fluctuation just before the bounce. The observed fluctuation arises because the kinetic energy contribution does not overwhelmingly dominate the system, unlike in the EKED regime. Consequently, the potential energy term gains significant relevance, a fact demonstrably evident in the second term of the expression (\ref{feg1}). Consequently, the energy density presents a spike in its behavior, which quickly dissipates, subsequently reaching its maximum  value $\rho_c$, which can be seen in bottom panel of figure \ref{fig:figs_KED}. In top right panel of Fig.~\ref{fig:figs_KED}, it is observed that the volume has an asymmetric behavior, in contrast to the EKED case, with a very pronounced drop before the bounce, after which its evolution relaxes. The increase in the noncommutative parameter is reflected in the growth of the volume relative to its commutative counterpart. The observed asymmetric behavior has been previously address in the existing literature. In \cite{Li:2018fco}, the model under consideration is a modification of the standard LQC, dubbed mLCQ-I, which takes a massless scalar field as its matter source. Meanwhile in \cite{Mohammadi:2024lgo}, a noncommutative extension of this mLCQ-I model, which includes a scalar potential, is address.

\begin{figure}[H]
\centering
  \includegraphics[scale=.4]{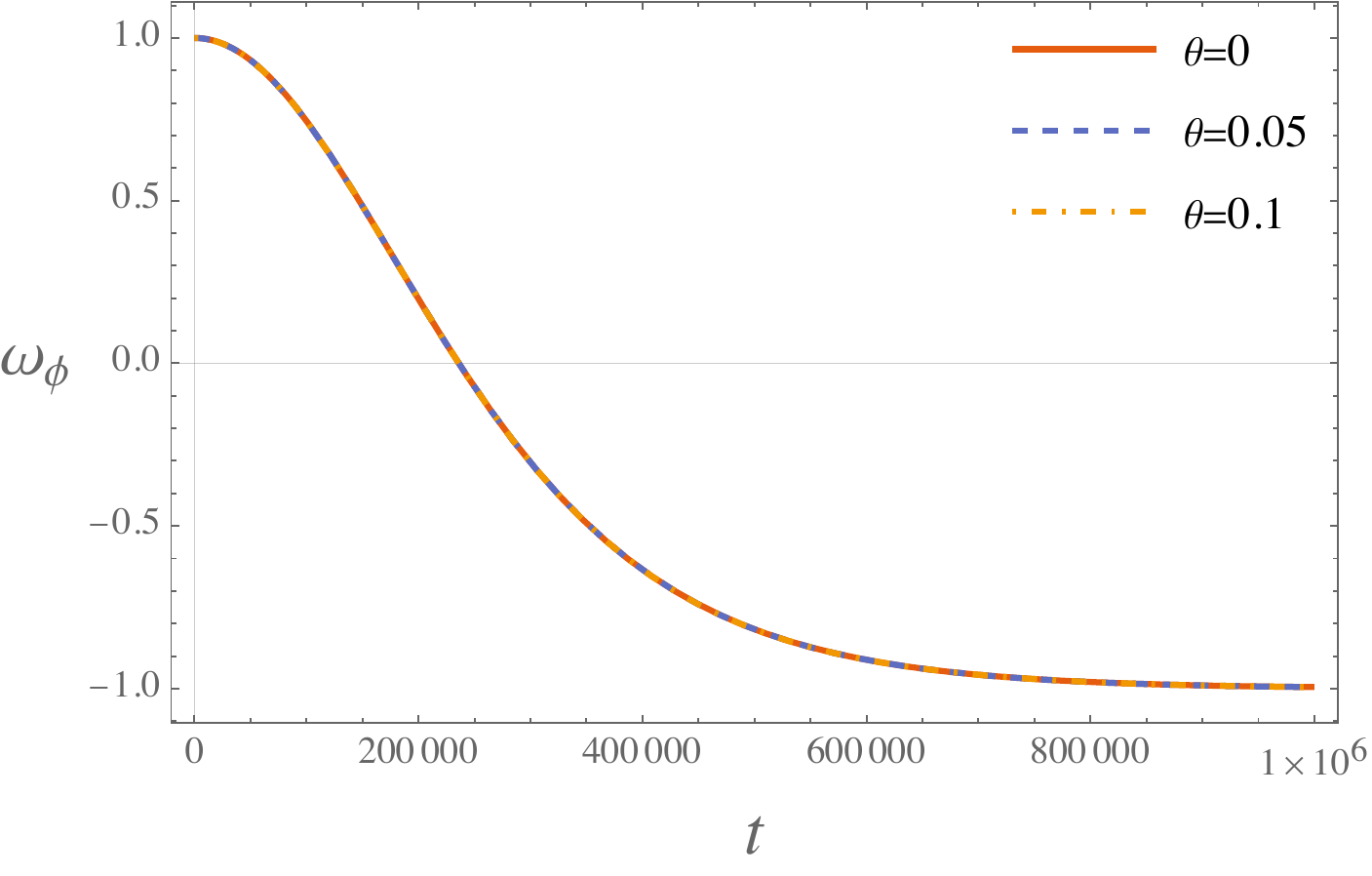}
  \caption{The figure shows the evolution of the effective equation of state parameter in the KED phase}
  \label{fig:omega_KED}
\end{figure}

As shown in figure \ref{fig:omega_KED}, the dynamics of $\omega_{\phi}$ (in the KED regime) reveals the different (pre)-inflationary stages of the universe, a bouncing stage, a transition period and a inflationary epoch (as in the EKED case). 
\subsubsection{Bouncing Phase}
\begin{figure}[H]
  \includegraphics[width=.47\linewidth]{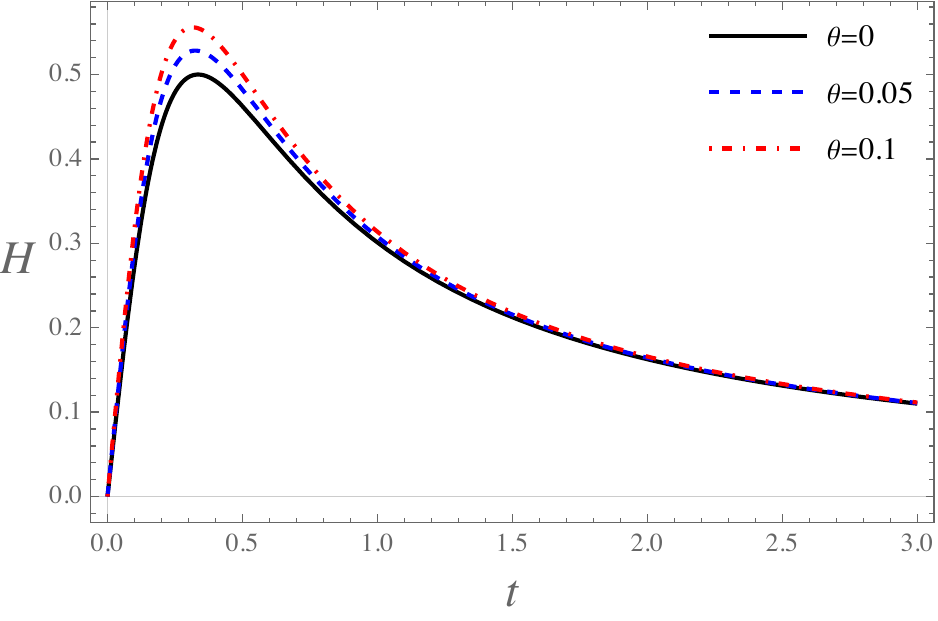}\hfill
   \includegraphics[width=.47\linewidth]{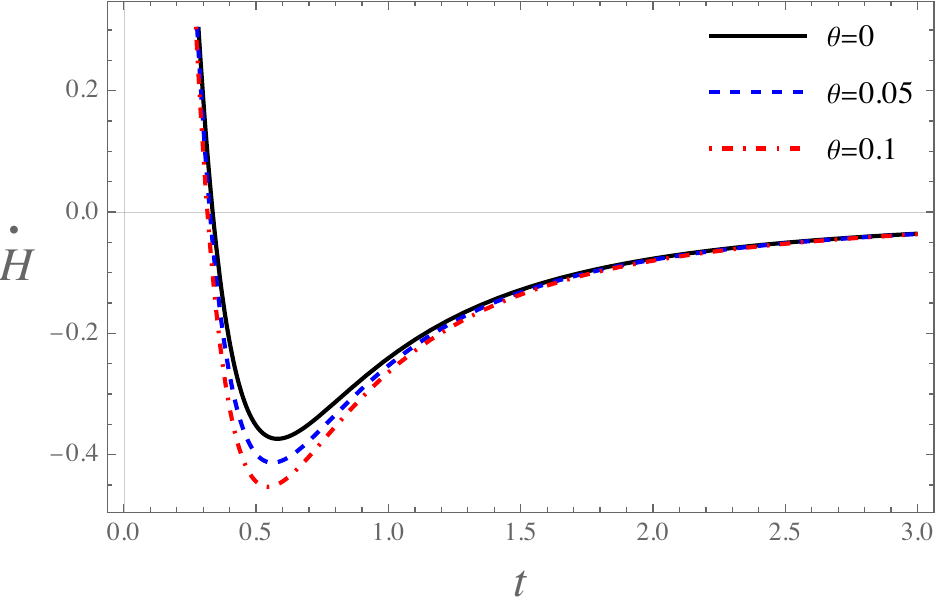}
 \caption{The figure illustrates the behavior of the Hubble parameter, as well as its corresponding rate of change, in the KED regime. In the left panel, we can observe that noncommutativity reduces the duration of the SI phase. The right panel shows that the SI phase concludes when $\dot H=0$.}
\label{fig:figs_KED1}
\end{figure}

As previously established, the bouncing phase initiates at the bounce and ceases once the once the effective EoS vanishes. We will denote these two moments as $t_{B}^{KED}$ and $t_{EQ}^{KED}$, respectively. Based on the numerical findings, summarized concisely in table \ref{tab:KED_Starobinsky}, it's evident that for the commutative scenario, $\theta=0$, the bounce phase concludes at $t_{EQ}^{KED}=2.394\times 10^5$. This phase also includes a period of super-inflation where the Hubble parameter undergoes a very rapid increase, achieving a maximum of $H=0.5$. The left panel of Fig.~\ref{fig:figs_KED1} shows the behavior of the Hubble parameter in the SI stage. From the right panel of figure \ref{fig:figs_KED1}, we can observe that the SI period ends when $\dot H=0$, that is, when $t_{SI}^{KED}=0.3357$ and the efolds evolve to be $N_{SI}^{KED}=0.1162$.
As shown in the right side of figure \ref{fig:figs_KED1}, the SI stage gets affected by noncommutativity, with the degree of this affection being determined by the scalar field evolution throughout this interval, leading to a reduction of the SI period, while simultaneously raising the maximum values of the Hubble parameter in comparison to the EKED regime. For $\theta=0.5$ the SI period ends when $t_{SI_{0.05}}^{KED}=0.3252$ and the maximum of the Hubble parameter is $H_{SI_{0.05}}^{KED}=0.5283$. For $\theta=0.1$, the SI phase concludes at $t_{SI_{0.1}}^{KED}=0.3158$ and the Hubble parameter achieves a maximum of $H_{SI_{0.1}}^{KED}=0.5558$. For the commutative case the bouncing phase comes to an end at $t_{EQ}^{KED}=2.394\times 10^5$, while the noncommutative counterparts end at $t_{EQ_{0.05}}^{KED}=2.390\times 10^5$ and $t_{EQ_{0.1}}^{KED}=2.387\times 10^5$, for $\theta=0.05$ and $\theta=0.1$, respectively.

\begin{table}[H]
			\scriptsize
			\caption{The table shows the numerical results of the effective NC-LQC with the Starobinsky potential for the KED phase. The Events: bounce, the end of SI (super-inflation), the equilibrium point KE=PE, the onset of slow-roll inflation, and the end of inflation, are compared for $\theta=0$, $\theta=0.05$ and $\theta=0.1$.}
				\label{tab:KED_Starobinsky}
			\begin{center}
			\resizebox{\textwidth}{!}{%
				\begin{tabular}{c   l  c  c  c  c  c c}
					\hline
					$\bf{b=0}$ & \bf Event & $\boldsymbol t$ & $\boldsymbol{\phi}$ & $\boldsymbol{\dot\phi}$ &$\bf H$ & $ \bf\dot H$ & $\bf N$ \\ \hline
					\hline
					& Bounce & 0 & -2.8 & 0.8  & $6.123\times10^{-17}$& 3& 0\\
					& End SI &0.3357 & -2.5631 & 0.5660& 0.5 &$-9.1479\times10^{-17}$ & 0.1162  \\
					$\theta=0$& KE=PE & $2.394\times 10^{5}$& 0.9543 &$9.799\times10^{-7}$ & $1.725\times10^{-6}$ & $-4.465\times10^{-12}$& 4.5381 \\
					& Onset Slow Roll & $1.301\times10^6$ & 1.1816 &$1.279\times10^{-8}$ & $1.235\times10^{-6}$ &$8.019\times10^{-20}$& 5.9073 \\  
					& End of Inflation & $7.887\times10^7$ &.1226 &$-1.697\times10^{-7}$ & $6.017\times10^{-7}$& $6.017\times10^{-13}$& 93.176\\ \hline
					& Bounce & 0 & -2.8 & 0.8  & $6.123\times10^{-17}$& 3 &0\\
					& End SI & 0.3252 & -2.5686 & 0.5771 & 0.5283 & $2.5417\times10^{-15}$ &  0.1188  \\
					$\theta=0.05$\ & KE=PE & $2.390\times10^{5}$ & 0.9814 & $9.893\times10^{-7}$ & $1.729\times10^{-6}$ & $-4.484\times10^{-12}$ & 4.5620\\
					& Onset Slow Roll & $1.331\times 10^6$ & 1.2112 &$1.557\times10^{-8}$ & $1.236\times10^{-6}$ &$1.248\times10^{-20}$& 5.9696 \\ 
					& End of Inflation & $8.873\times10^7$ &.1226 &$-1.6973\times10^{-7}$ & $6.017\times10^{-7}$& $-3.620\times10^{-13}$& 105.334\\ \hline
					& Bounce & 0 & -2.8 & 0.8  & $6.123\times10^{-17}$& 3&0\\
					& End SI & 0.3158 & -2.5737 & 0.5871 & 0.5558 & $-2.0640\times10^{-16}$& 0.1213  \\
					$\theta=0.1$\ & KE=PE & $2.387\times10^5$ & 1.0051 & $9.974\times10^{-7}$ & $1.732\times10^{-6}$ &$-4.499\times10^{-12}$& 4.5843 \\
					& Onset Slow Roll & $1.356\times10^6$ & 1.2371 &$1.057\times10^{-8}$ & $1.237\times10^{-6}$ &$1.156\times10^{-19}$& 6.0256 \\ 
					& End of Inflation & $ ---$ & $---$ & $ ---$ & $ ---$ & $ ---$ & $---$\\ \hline
				\end{tabular}
				}
			\end{center}
		\end{table}

\subsubsection{Transition Phase}

The transition phase begins when the equation of state parameter $\omega_{\phi}=0$, this occurs when the kinetic energy is equivalent to the potential energy, and concludes when $\omega_{\phi}=-1$, as depicted in Fig.~\ref{fig:omega_KED}. For the commutative case this stage starts at $t_{EQ}^{KED}=2.394\times 10^5$ and ends at $t_{i}^{KED}=1.301\times 10^6$. For the noncommutative cases $\theta=0.05$ and $\theta=0.1$, the transition stage starts when $t_{EQ_{0.05}}^{KED}=2.390\times 10^5$ and $t_{EQ_{0.1}}^{KED}=2.387\times 10^5$, respectively; and terminates when $t_{i_{0.05}}^{KED}=1.331\times 10^6$ and $t_{i_{0.1}}^{KED}=1.356\times 10^6$, respectively. From this, we can observed that the noncommutative effects results in an extension of the duration of this stage. This characteristic is also evident in the evolution of the efolds. For the commutative case we have that $N_{tr}^{KED}=5.9073$, whereas, the noncommutative cases results in $N_{tr_{0.05}}^{KED}=5.9696$ and $N_{tr_{0.1}}^{KED}=6.0256$, for $\theta=0.05$ and $\theta=0.1$, respectively (see table \ref{tab:KED_Starobinsky}).
\subsubsection{Slow-Roll Inflation}

\begin{figure}[H]
  \includegraphics[width=.47\linewidth]{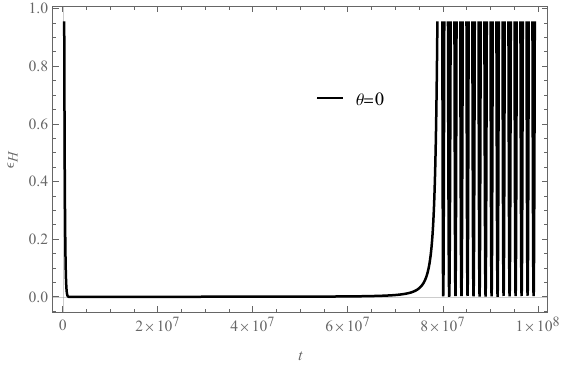}\hfill
   \includegraphics[width=.47\linewidth]{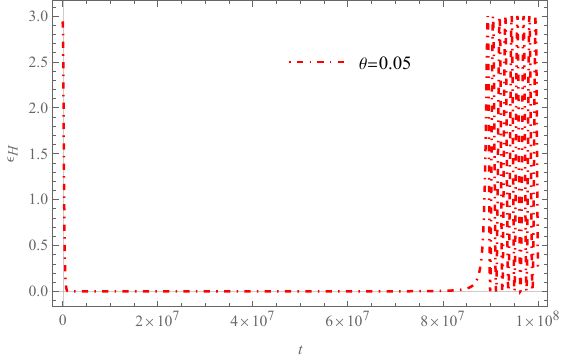}
   \centering
   \includegraphics[width=.47\linewidth]{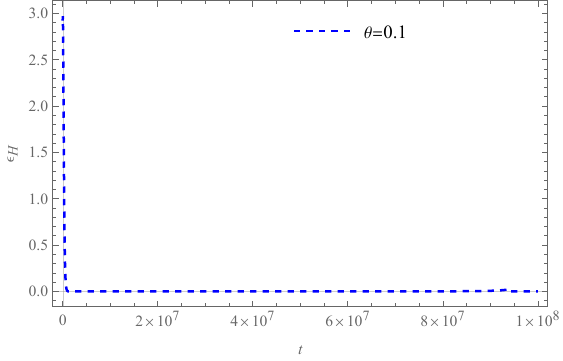}
 \caption{The figure illustrates the behavior of the slow-roll parameter $\epsilon_H$ in the KED regime. The initial conditions for the numerical solutions are $\phi=-2.8~m_{pl}$, $\dot\phi=0.8~m_{pl}^2$.}
\label{fig:figs_epsilonKED}
\end{figure}

For the slow-roll phase to be realized $\omega_{\phi}=-1$, where the energy density is fully govern by the Starobinsky potential. From table \ref{tab:KED_Starobinsky} we can extract that this stage commences at $t_{i}^{KED}=1.301\times 10^6$ for the commutative case, while for the noncommutative counterparts we have $t_{i_{0.05}}^{KED}=1.331\times 10^6$ and $t_{i_{0.1}}^{KED}=1.356\times 10^6$, for $\theta=0.05$ and $\theta=0.1$, respectively. For the commutative scenario, the inflationary period concludes at $t_{sr}^{KED}=7.887\times10^7$, see the top left panel of figure~\ref{fig:figs_epsilonKED}. The introduction of noncommutative effects modifies the end of the inflationary period. Specifically, for a noncommutative parameter $\theta=0.05$ the end-time is shifted to $t_{sr_{0.05}}^{KED}=8.873\times10^7$, see the top right panel of figure~\ref{fig:figs_epsilonKED}. Furthermore, for $\theta=0.1$ the period terminates at $t_{sr_{0.1}}^{KED}=1\times10^8$, where the slow-roll parameter $\epsilon_H\to0$ implying that, given the initial conditions $\phi= -2.8~m_{pl}$, $\dot\phi=0.8~m_{pl}^2$, inflation does not appear to have an end up to $t=1\times10^8$, as demonstrated by the behavior of $\epsilon_H$, see Fig~\ref{fig:figs_epsilonKED} .

\begin{figure}[H]
\centering
  \includegraphics[scale=1]{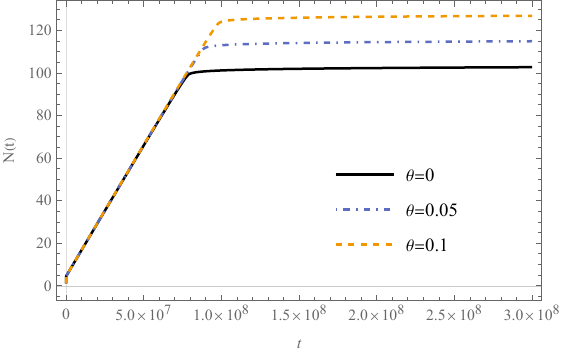}
  \caption{Evolution of the e-folds in the kinetic energy domination phase. We can see that as the noncommutative parameter increases so does the number of e-folds.}
  \label{fig:efolds_KED}
\end{figure}

%
The number of e-folds aligns with the observational data \cite{Planck:2018jri}, which establishes a bound of $N\geq 60$.  Additionally, from the enlarged section of Fig.~\ref{fig:efolds_KED}, we can see that as the noncommutative parameter increases, so does the number of e-folds.  

\subsection{PE Domination: After the Bounce}
\begin{figure}[H]
  \centering
   \includegraphics[width=0.8\linewidth]{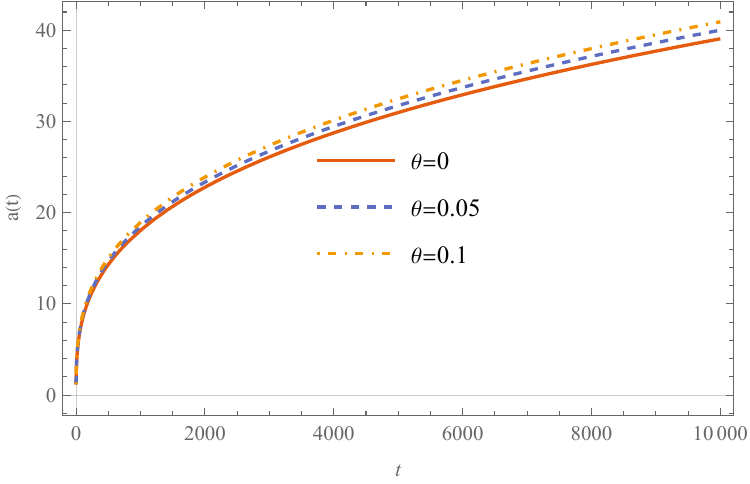}
  \caption{The figure shows the evolution of the scale factor in the potential energy domination regime. We can see that the scale factor grows as $\theta$ increases.}
  \label{fig:New_scale_factor_PED}
 \end{figure}

In this stage, the potential energy dominates the evolution of the universe after the bounce, and the ratio of the potential energy at the bounce to critical density is $F_B>0.5$. Here we take the initial conditions $\phi_B=-3.23161~m_{pl}$, $\dot\phi_B=0.8~m_{pl}^2$. The scale factor at this stage has a slower growth compared to EKED and KED, see figure \ref{fig:New_scale_factor_PED}, where the noncommutative corrections causes the scale factor to increase. 
\subsubsection{Bouncing Phase}
\begin{figure}[H]
  \includegraphics[width=0.5\linewidth]{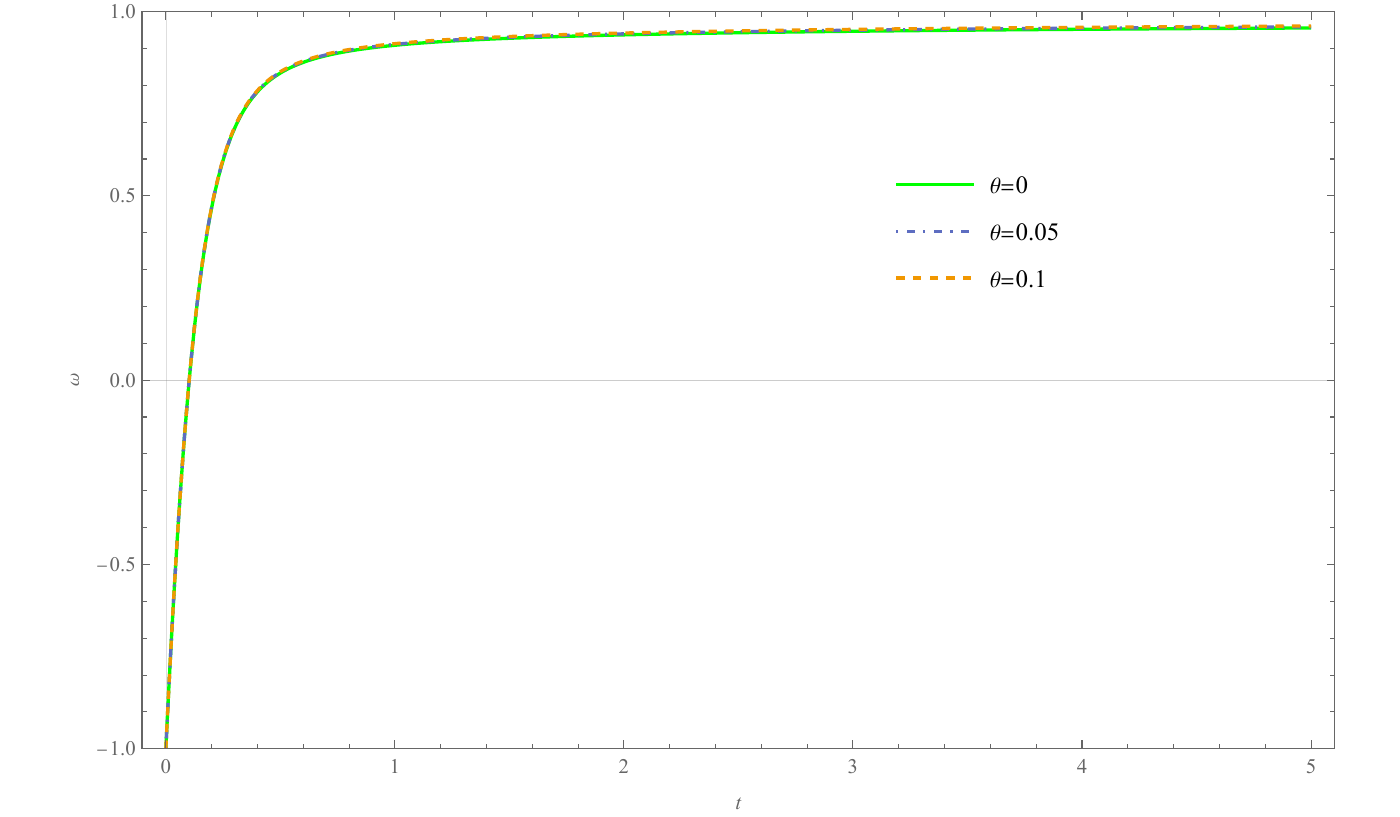}\hfill
  \includegraphics[width=0.5\linewidth]{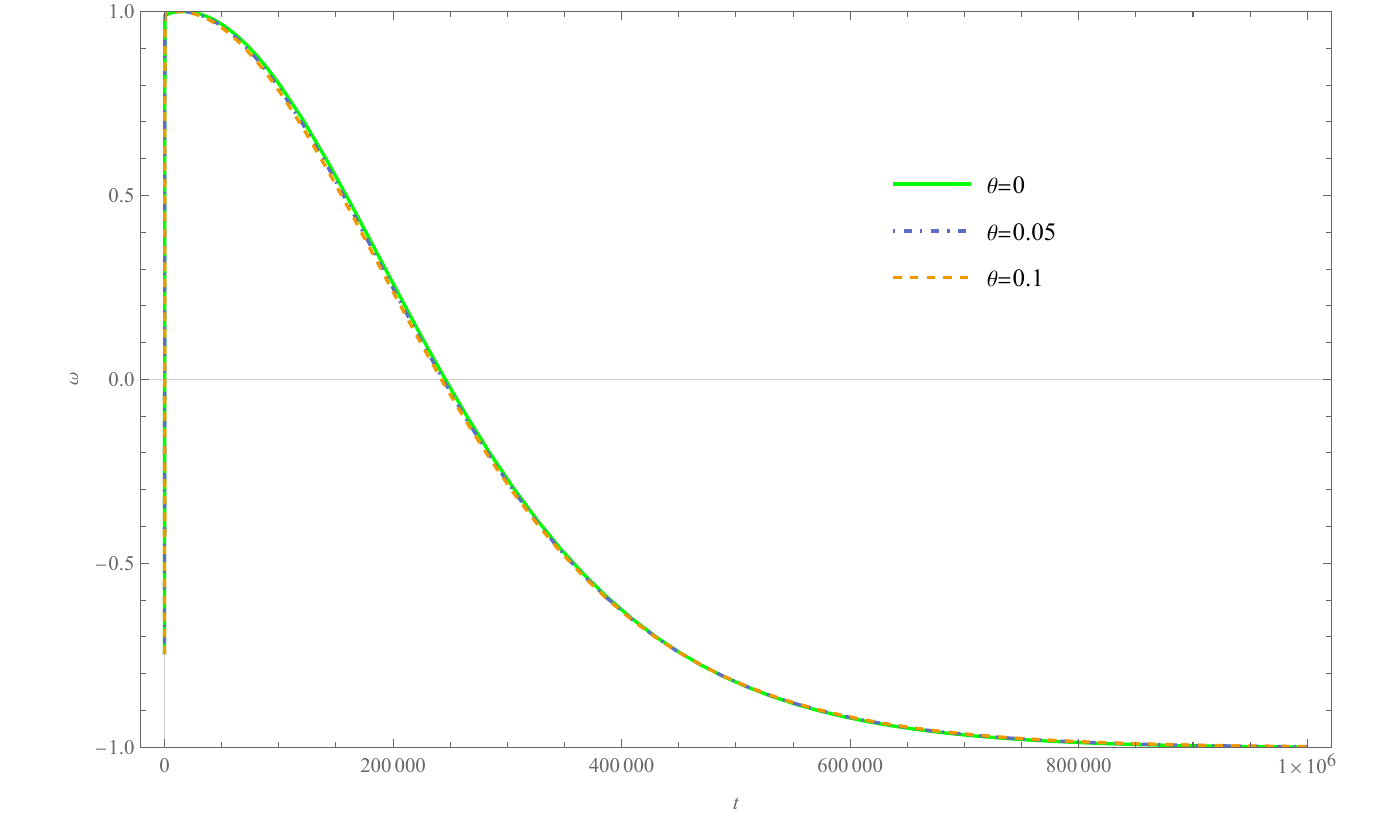}
  \centering
  \includegraphics[width=0.5\linewidth]{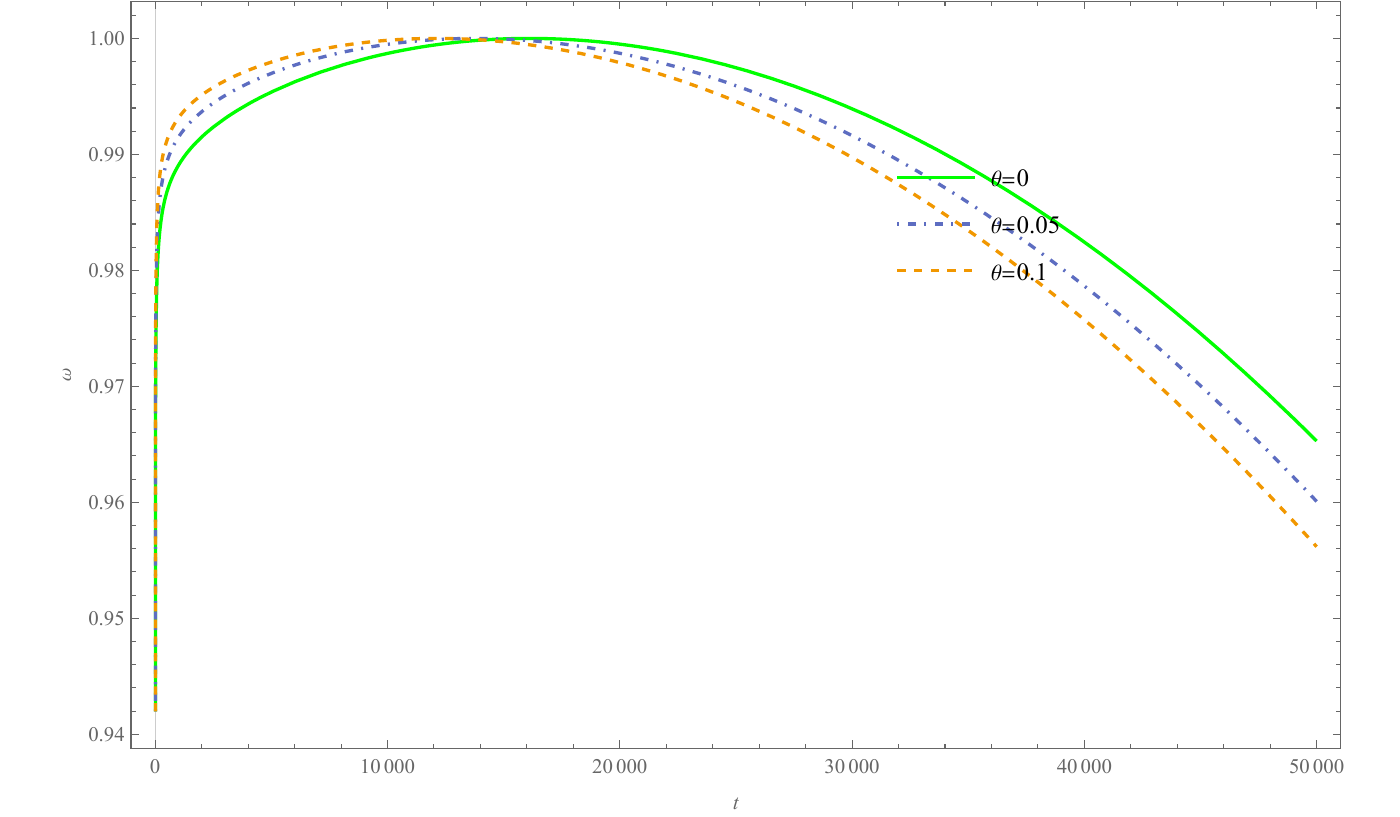}
   \caption{The figure shows the behavior of the effective EoS parameter in the potential energy dominated regime.}
\label{fig:figs_omegaPED}
\end{figure}

\begin{figure}[H]
  \includegraphics[width=0.5\linewidth]{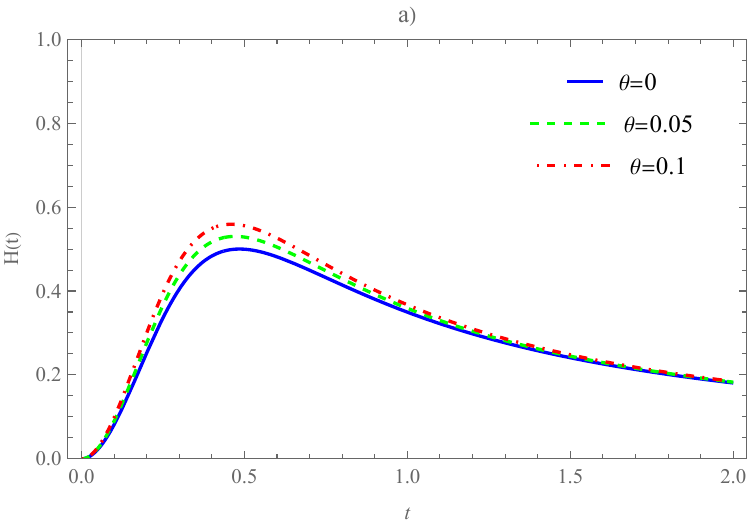}\hfill
  \includegraphics[width=0.5\linewidth]{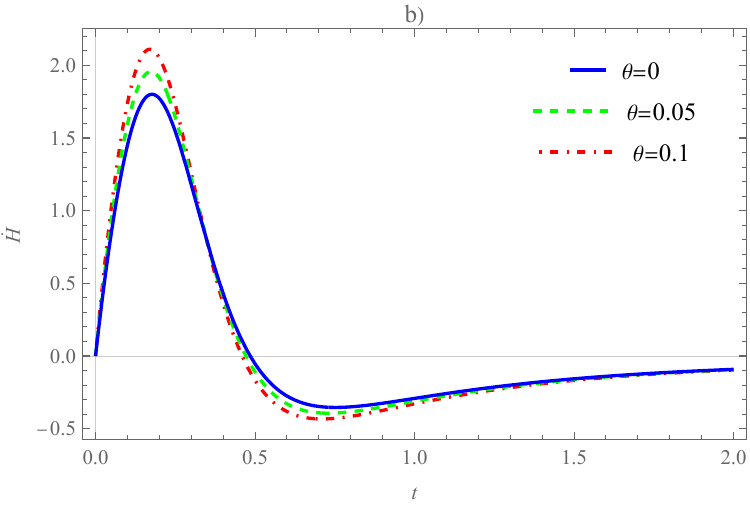}
   \caption{The figure shows the behavior of the Hubble parameter (left panel) and its rate of change (right panel), in the potential energy dominated regime.}
\label{fig:figs_HPED}
\end{figure}

The evolutionary dynamics of the effective equation of state parameter, as depicted in the upper left panel of Figure~\ref{fig:figs_omegaPED}, reveal a critical transition during the bounce phase. Specifically, attains a value of $-1$ at the bounce and subsequently increases, reaching an asymptotic value of $1$ (where $t=16107.8$, see bottom panel of Fig.~\ref{fig:figs_omegaPED}). Following the bounce, there is an epoch of accelerated expansion in the SI stage, whose duration is $t=0.177065$. $H(t)$ begins to decelerate until reaching its maximum value of 0.5, as in the EKED and KED cases, this behavior is shown in the right panel of Fig.~\ref{fig:figs_HPED}. The noncommutative counterpart of $H(t)$ in this period has the same behavior as for the KED regime, as the value of $\theta$ increases the Hubble parameter has greater amplitude, as shown in Fig.~\ref{fig:figs_HPED}. From table~\ref{tab:PED_Starobinsky} we can extract that the SI stage ends at $t_{SI}^{PED}=0.4850$, this means that when the potential energy dominates at the bounce, the duration of the SI period is increased, compared to the EKED and KED, that is $t_{SI}^{EKED}<t_{SI}^{KED}<t_{SI}^{PED}$.
Upon attaining the value $\omega_{\phi}=1$, the dynamic evolution of the system exhibits a congruence with the characteristics observed within the EKED and KED regimes. As is evident in the lower portion of Fig.~\ref{fig:figs_omegaPED}, the noncommutative models undergo a behavioral transition during the bouncing phase upon reaching $\omega_{\phi}=1$. Prior to this value, an increase in the noncommutative parameter correlates directly with an enhanced growth rate of $\omega_{\phi}$ as previously established. Conversely, subsequent to attaining the maximum value, this dependency is inverted: larger values of $\theta$ correspond to diminished values of $\omega_{\phi}$. This reversal in scaling behavior aligns precisely with the domains previously subjected to analysis. From table~\ref{tab:PED_Starobinsky}, we can see that the SI phase ends at $t_{EQ}^{PED}=2.246\times10^5$.


\begin{table}[H]
			\scriptsize
			\caption{The table shows the numerical results of the effective NC-LQC with the Starobinsky potential for the PED phase. The Events: bounce, the end of SI (super-inflation), the equilibrium point KE=PE, the onset of slow-roll inflation, and the end of inflation, are compared for $\theta=0$, $\theta=0.05$ and $\theta=0.1$.}
				\label{tab:PED_Starobinsky}
		
			\begin{center}
			\resizebox{\textwidth}{!}{%
				\begin{tabular}{c   l  c  c  c  c  c c}
					\hline
					$\bf{b=0}$ & \bf Event & $\boldsymbol t$ & $\boldsymbol{\phi}$ & $\boldsymbol{\dot\phi}$ &$\bf H$ & $ \bf\dot H$ & $\bf N$ \\ \hline
					\hline
					& Bounce & 0 & -3.3216118 & 0.8  & $6.123\times10^{-17}$&$ -4.441\times10^{-15}$& 0\\
					& End SI &0.4850 & -2.9377 & 0.6213& 0.5 &$8.29024\times10^{-16}$ & 0.139605  \\
					$\theta=0$& KE=PE & $2.246\times 10^{5}$& 0.7119 &$9.648\times10^{-7}$ & $1.6652\times10^{-6}$ & $-4.159\times10^{-12}$& 4.6159 \\
					& Onset Slow Roll & $1.028\times10^6$ & 0.9302 &$3.824\times10^{-8}$ & $1.217\times10^{-6}$ &$1.845\times10^{-19}$& 5.6187 \\  
					& End of Inflation & $2.963\times10^7$ &0.12263 &$-1.6973\times10^{-7}$ & $6.01677\times10^{-7}$& $-3.62015\times10^{-13}$& 32.9665\\ \hline
					& Bounce & 0 & -3.3216118 & 0.8  & $6.771\times10^{-17}$& $-4.912\times^{-15}$ &0\\
					& End SI & 0.4715 & -2.9447 & 6378 & 0.530121 & $1.99785\times10^{-16}$ &  0.14345  \\
					$\theta=0.05$\ & KE=PE & $2.436\times10^{5}$ & 0.7650 & $9.903\times10^{-7}$ & $1.6838\times10^{-6}$ & $-4.253\times10^{-12}$ & 4.6380\\
					& Onset Slow Roll & $1.085\times 10^6$ & 0.9918 &$3.119\times10^{-8}$ & $1.223\times10^{-6}$ &$-3.108\times10^{-19}$& 5.7201 \\ 
					& End of Inflation & $3.744\times10^7$ &.0.122603 &$-1.6973\times10^{-7}$ & $6.01678\times10^{-7}$& $-3.62016\times10^{-13}$& 42.4349\\ \hline
					& Bounce & 0 & -3.3216118 & 0.8  & $7.418\times10^{-17}$& $-5.383\times10^{-15}$&0\\
					& End SI & 0.4596 & -2.9511 & 0.6524 & 0.559056 & $1.48953\times10^{-16}$& 0.147063  \\
					$\theta=0.1$\ & KE=PE & $2.422\times10^5$ & 0.8103 & $1.010\times10^{-6}$ & $1.6968\times10^{-6}$ &$-4.319\times10^{-12}$& 4.6596 \\
					& Onset Slow Roll & $1.135\times10^6$ & 1.0420 &$2.618\times10^{-8}$ & $1.227\times10^{-6}$ &$-6.721\times10^{-20}$& 5.8090 \\ 
					& End of Inflation & $ 4.569\times10^{7}$ & $0.122604$ & $ -1.69731\times10^{-7}$ & $ 6.01670\time10^{-7}$ & $ -3.62018\times10^{-13}$ & $52.4958$\\ \hline
				\end{tabular}
				}
			\end{center}
		\end{table}


\subsubsection{Transition Phase}

In the potential energy dominated regime, the transition phase lasts at $N=5.6187$ after the bounce occurs (see table~\ref{tab:PED_Starobinsky}). Observation of the right panel of Fig.~\ref{fig:figs_omegaPED} reveals that the influence of noncommutativity on the evolution of the effective EoS parameter, is more pronounced than in the preceding scenarios. Consistent with earlier trends, an increase in the noncommutative parameter $\theta$ correlates with a more rapid decrease in the effective EoS parameter. This characteristic persists until the model with $\theta=0.05$ reaches the time $t=471651$. At this juncture, the noncommutative values of $\omega_{\phi}$ begin to increase relative to the commutative counterpart. A similar behavior is observed for the $\theta=0.1$ model, where this shift occurs at the earlier time $t=468701$. This change is quantitatively substantiated in table~\ref{tab:PED_Starobinsky}, where a comparison of the end of the transition phases demonstrates the relationship: $N^{PED}_{tr}< N^{PED}_{tr_{0.05}}<N^{PED}_{tr_{0.1}}$.

It can be observed from the upper right panel of Fig.~\ref{fig:figs_omegaPED} that the effects of noncommutativity on the evolution of the effective EoS parameter are more noticeable than in the previous cases, and following the same behavior, the larger $\theta$ is, the faster the EoS parameter decreases. This remains until the model for $\theta=0.05$ reaches $t=471651$ where now the non-commutative $\omega_\phi$ values increase compared to its commutative counterpart. This behavior is similar for the $\theta=0.1$ model, where now $t=468701$. In Table~\ref{tab:PED_Starobinsky} we can see this change by comparing the ends of the transition phases $N^{PED}_{tr}< N^{PED}_{tr_{0.05}}<N^{PED}_{tr_{0.1}}$.

\subsubsection{Slow-roll Inflation}
\begin{figure}[H]
  \includegraphics[width=.47\linewidth]{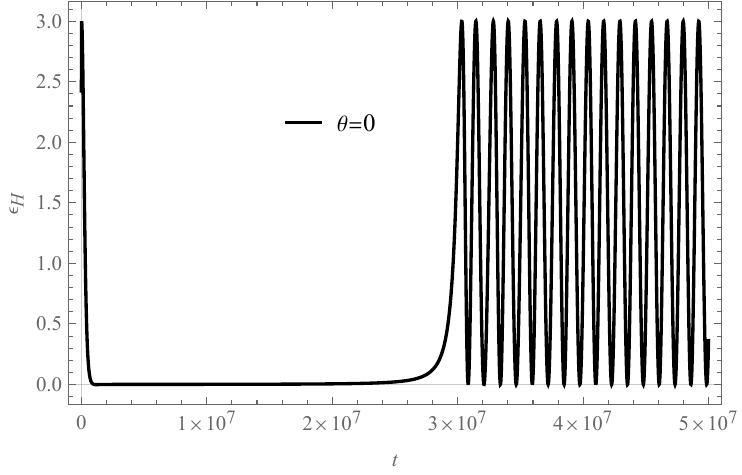}\hfill
   \includegraphics[width=.47\linewidth]{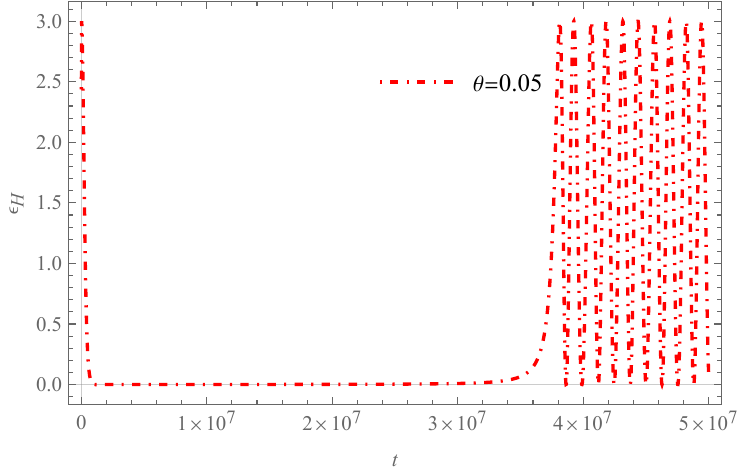}
   \centering
   \includegraphics[width=.47\linewidth]{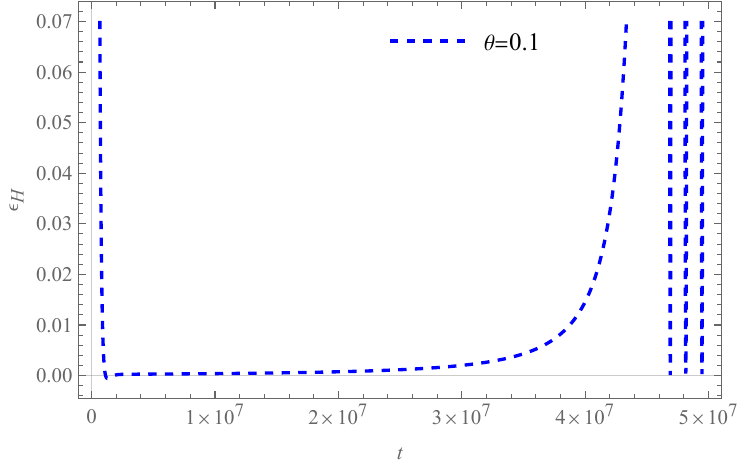}
 \caption{The figure illustrates the behavior of the slow-roll parameter $\epsilon_H$ in the PED regime. The initial conditions for the numerical solutions are $\phi=-3.32161~m_{pl}$, $\dot\phi=0.8~m_{pl}^2$.}
\label{fig:figs_epsilonPED}
\end{figure}

\begin{figure}[H]
\centering
  \includegraphics[width=.75\linewidth]{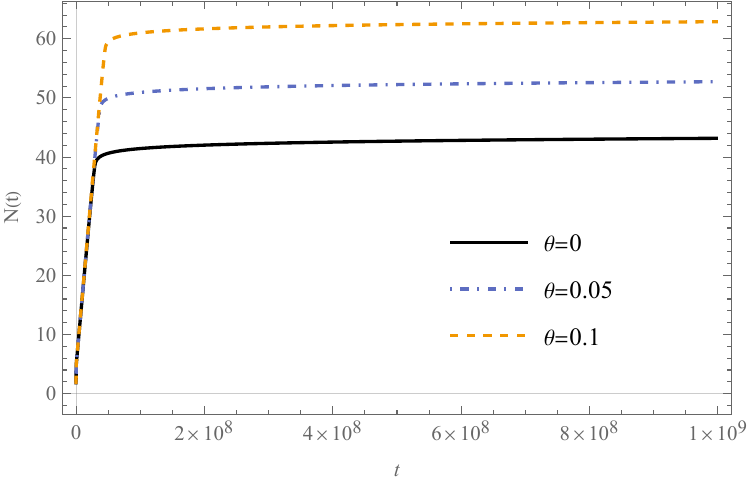}
 \caption{Numerical solutions with the initial conditions are $\phi=-3.32161~m_{pl}$, $\dot\phi=0.8~m_{pl}^2$. The figure illustrates the behavior of the evolution of the e-folds in the PED regime}
\label{fig:figs_efolds_PED}
\end{figure}

Fig.~\ref{fig:figs_epsilonPED} illustrates the evolution of the slow-roll parameter, $\epsilon_H$, which confirms the presence of an inflationary period in both the commutative and noncommutative cosmological models under consideration. It is discernible that the noncommutative parameter $\theta$ exhibits a similar qualitative effect to that observed in the KED model. Consistent with the findings of the preceding section, $\theta$ exerts a more pronounced influence on the PED model, specifically by preventing eternal inflation for the case where $\theta=0.1$ (as shown in the lower panel of figure~\ref{fig:figs_epsilonPED} ). For both the commutative model and the noncommutative model with $\theta=0.05$, figure~\ref{fig:figs_efolds_PED} indicates a number of e-folds, is less than the requisite $N<60$. Consequently, the duration of the inflationary period in these scenarios does not satisfy the necessary cosmological constraint reported in reference \cite{Planck:2018jri}. The model with $\theta=0.1$ appears to reach the minimum e-fold bound; however, the e-fold number displayed in Fig.~\ref{fig:figs_efolds_PED} is calculated from the time of the cosmological bounce, not from the onset of the slow-roll phase. When the number of e-folds is computed precisely over the duration of the slow-roll phase, which commences at $t^{PED}_i=1.135\times10^6$ and terminates at $t^{PED}_{sr}=4.569\times10^{7}$, the corresponding value is $N^{PED}_{sr}=52.4958$. This value again fails to satisfy the minimum requirement of $N\geq60$. Evidently, the noncommutativity effect dictates that an increase in $\theta$ correlates with a greater number of e-folds. While this suggests the possibility to achieve the requisite e-folds for a viable inflationary period, it requires $\theta$ to be of the order $\theta\sim 1$. Such a significant noncommutativity parameter should, therefore, be observationally apparent.


This is consistent with the results presented in \cite{Bonga:2015kaa,Li:2019ipm}, where one of the main conclusions establishes that when the potential energy dominates at the bounce it is not possible for the universe to enter a phenomenologically viable inflationary stage. 

\section{Evolution of the Scalar Field: A Dynamical System Approach}\label{dynamical_system}
\begin{figure}[H]
  \includegraphics[width=0.45\linewidth]{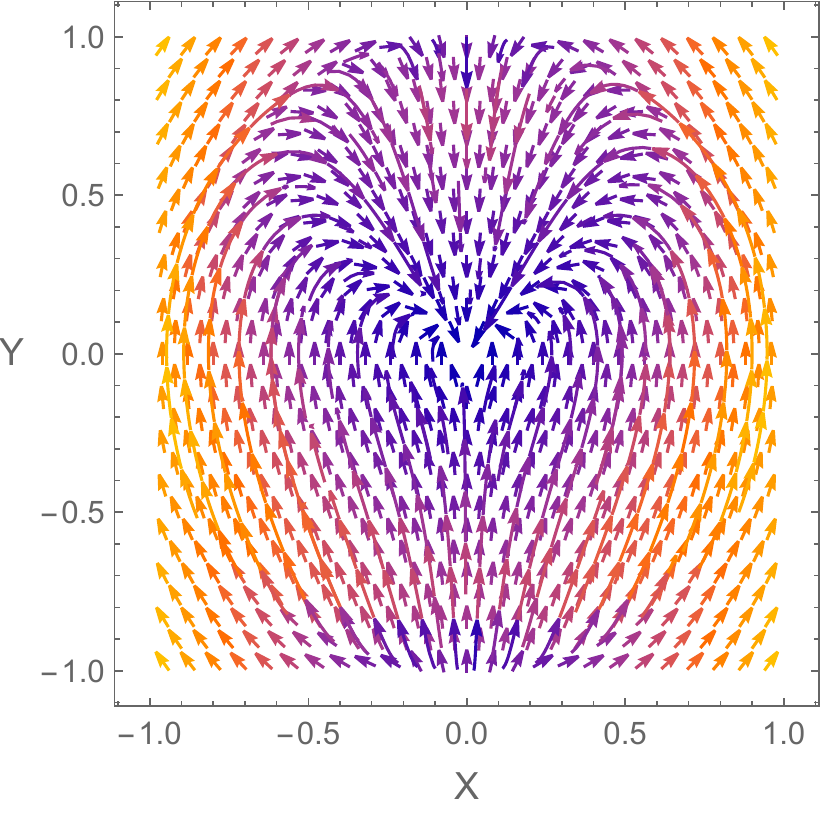}\hfill
  \includegraphics[width=0.50\linewidth]{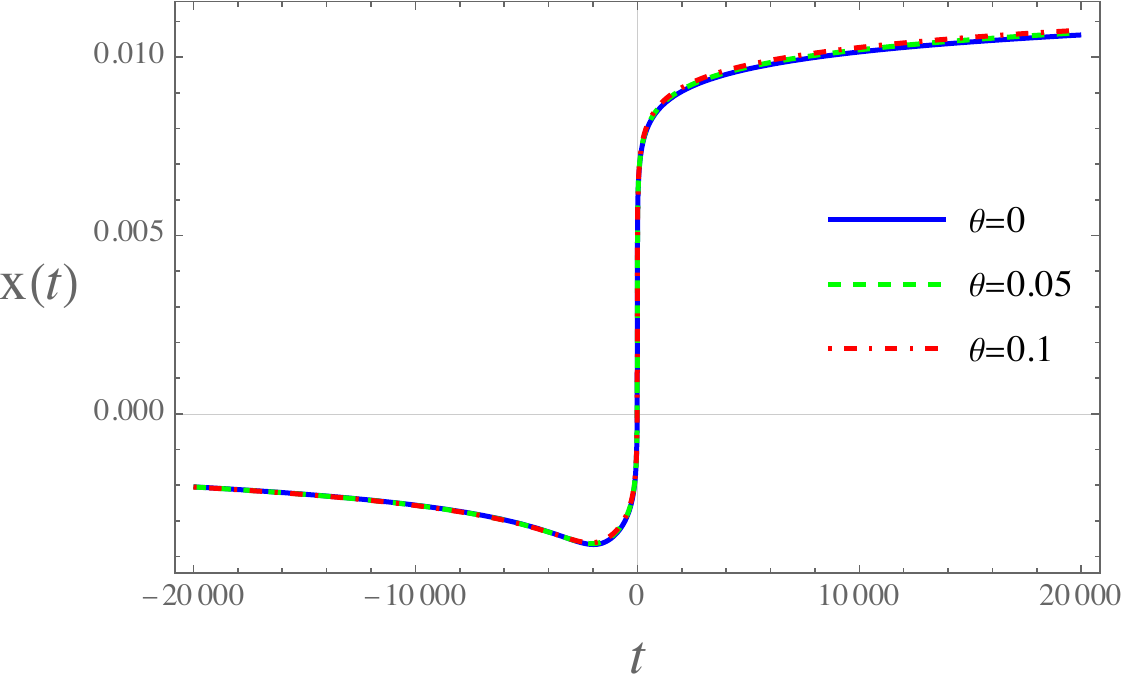}
   \caption{The figure shows the behavior of the inflaton field in phase space (left panel) and the trajectory of the scalar field (right panel).}
   \label{fig:figs_atractores}
\end{figure}

This section provides a brief overview of the general behavior of the scalar field from the perspective of dynamical systems. Through the definition of the new set variables 
\begin{equation}
    x=\sqrt{\frac{3m^2}{32\pi G \rho_c}}\left(1-e^{-\sqrt{\frac{16\pi G}{3}}\phi}\right),\qquad y=\frac{\dot\phi}{\sqrt{2\rho_c}},
    \label{eqdinsys}
\end{equation}
where its domain satisfies the condition $x^2+y^2=\rho/\rho_c$, the field equations related to (\ref{ham_eff_nc}) can be reformulated as an autonomous system as
\begin{eqnarray}
    \dot x&=&m\left(1-\sqrt{\frac{32\pi G\rho_c}{3m^2}}x\right)y,\\
    \dot y&=&-\frac{3y}{\gamma\lambda}\left(x^2+y^2\right)^{1/2}\left(1-x^2-y^2\right)^{1/2}-mx\left(1-\sqrt{\frac{32\pi G\rho_c}{3m^2}}x\right ).
\end{eqnarray}
As indicated by the governing equation above, there is no explicit dependence on the noncommutative parameter; consequently, the phase space behavior of the inflaton field is depicted in the left panel of Figure~\ref{fig:figs_atractores}.
This behavior is consistent with the one presented in \cite{Li:2018fco}. In the right panel of figure \ref{fig:figs_atractores} the trajectory of the scalar field is shown. We can observed the super inflation phase (vertical line), as well as the slow-roll phases at the decreasing/increasing extremes of the plot.

\begin{figure}[H]
\centering
  \includegraphics[width=0.45\linewidth]{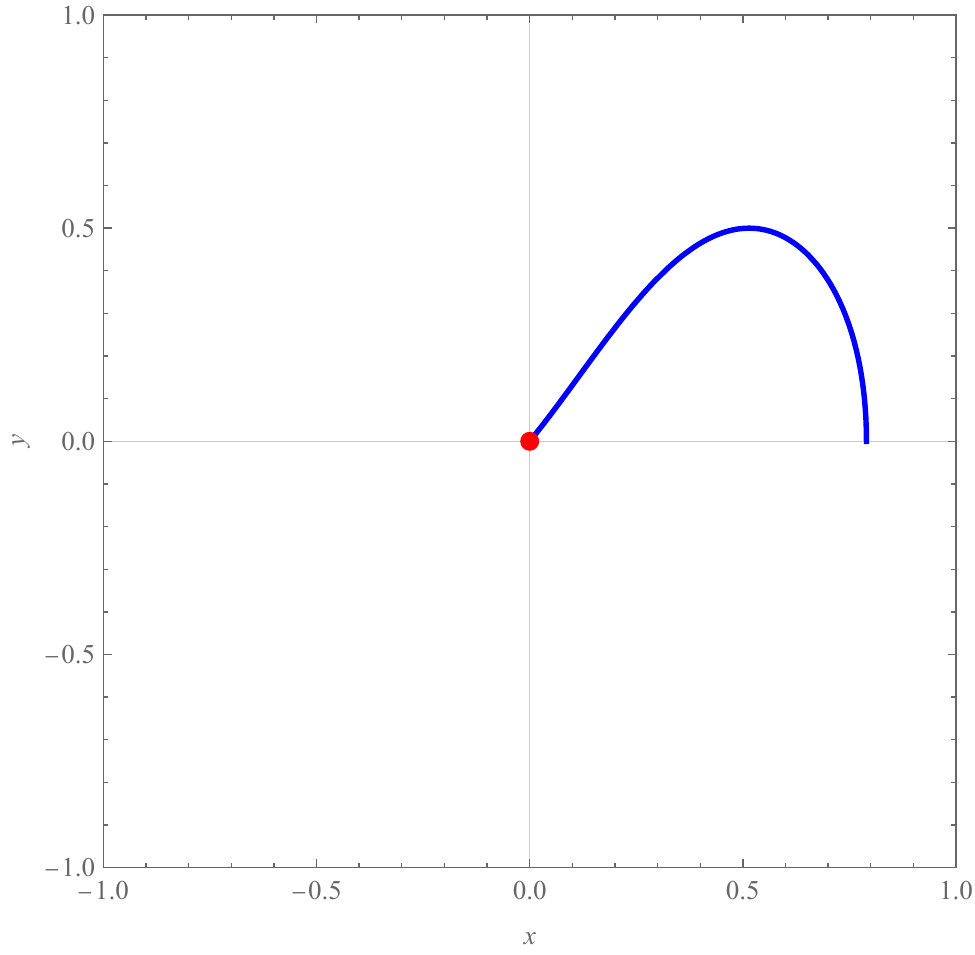}
   \caption{In the figure, we can see that the trajectories remain in proximity to the origin, thereby confirming the stability of the fixed point at $(0, 0)$.}
   \label{fig:figs_atractores1}
\end{figure}

This dynamical system admits fixed points at $(0,0)$ and $\left(\frac{1}{2}m,0\right)$ within the domain of interest. Upon identification of these fixed points, their stability properties are examined by analyzing the behavior of neighboring trajectories. Specifically, the eigenvalues of the Jacobian matrix $J = D(\vec{f}(\vec{x}))$, evaluated at the fixed point, are computed. At $(0,0)$, the resulting eigenvalues are $\pm i2^{1/4}m$, both of which are purely imaginary. This indicates that the fixed point is a linear center, with nearby trajectories forming closed or quasi-periodic orbits. Owing to the non-hyperbolic nature of these eigenvalues (i.e., vanishing real part), linearization alone is insufficient for establishing global stability. Accordingly, the system is further analyzed in polar coordinates to rigorously ascertain its stability properties. Fig.~\ref{fig:figs_atractores1} demonstrates that the trajectories remain in proximity to the origin, thereby confirming the stability of the fixed point at $(0,0)$.
For the second fixed point, $(\frac{1}{2}m, 0)$, the associated eigenvalues are $\{0, \frac{3}{4}m\sqrt{4 - m^2}\}$. The presence of a zero eigenvalue signals that this fixed point is non-hyperbolic. Additionally, for the second eigenvalue, noting that $m^2 \ll 4$, its magnitude is small and positive, indicating an attracting direction. However, due to the non-hyperbolic nature of the fixed point, linear stability analysis is inconclusive, necessitating a more comprehensive nonlinear stability investigation. 

\begin{figure}[H]
\centering
  \includegraphics[width=0.45\linewidth]{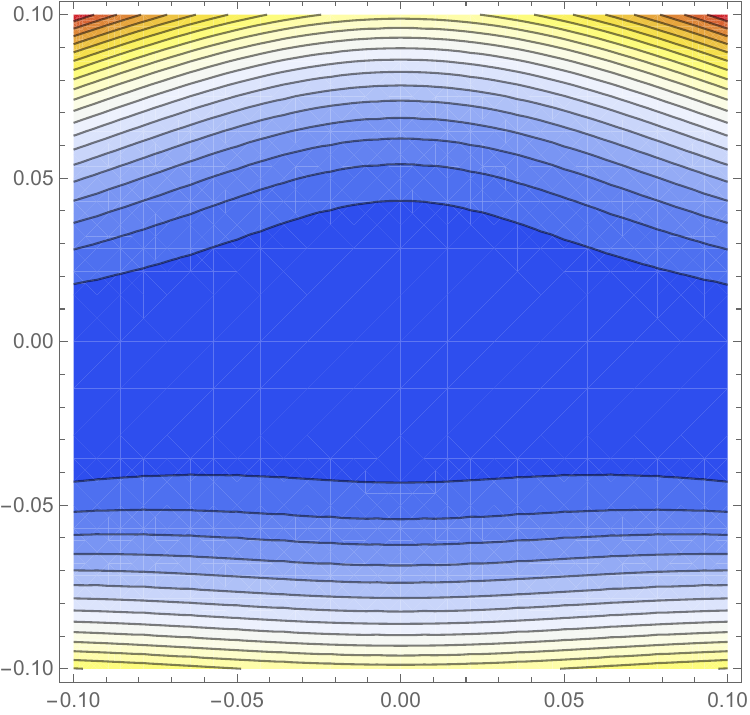}
  \includegraphics[width=0.10\linewidth]{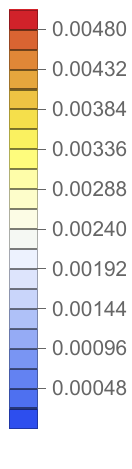}
   \caption{The figure illustrates that in the vicinity of the fixed point $\dot{L}(x, y) > 0$, indicating that the point is $(\frac{m}{2}, 0)$ is unstable.}
   \label{fig:figs_atractores3}
\end{figure}

To facilitate a nonlinear stability analysis, we construct a Lyapunov function. A natural choice for such a function is one that quantifies the quadratic distance from the fixed point, expressed as
\begin{equation}
\label{Lyapunovfunc}
L(x,y)=\frac{1}{2}(x-x^*)^2+\frac{1}{2}y^2,
\end{equation}
taking that the fixed point is $(x^*,0)$. To assess the stability of the fixed point, we examine the sign of $\dot{L}(x, y)$. As illustrated in Fig.~\ref{fig:figs_atractores3}, an analysis in the vicinity of the fixed point reveals that $\dot{L}(x, y) > 0$, indicating that the point $(\frac{m}{2}, 0)$ is unstable.

\section{Discussion: Starobinsky Potential vs. Quadratic Potential}\label{discusion}
In this section we address a comparative analysis of results obtained for the quadratic scalar field potential \cite{Diaz-Barron:2023ctp} and the ones developed in this work for the Starobinsky potential, for the three energy regimes, all within the framework of the NC ELQC formalism. 

\subsection{Starobinsky potential vs. Quadratic potential:  EKED Regime}
A primary distinction observed is that noncommutativity extends the duration of various cosmological phase, namely, the bounce, transition, and slow-roll inflation for the quadratic potential. Conversely, the Starobinsky potential exhibits the opposite behavior.

During the SI period, the results for both models are highly similar in the commutative case. However, this similarity does not extend to the noncommutative scenarios. For instance, the super-inflation phase concludes earlier for the Starobinsky potential compared to the quadratic potential. Regarding the scalar field in the $\phi^2$ model, its value decreases as the noncommutative parameter increases, a trend reversed in the Starobinsky potential. For the Hubble parameter, an increase is observed in the quadratic potential model as $\theta$ grows, whereas the Starobinsky potential demonstrates the opposite trend. This behavior is also evident in the evolution of e-folds.

The differences become more pronounced during the transition period. For example, in the $\phi^2$ model, the duration of this period lengthens with increasing noncommutative parameter. In contrast, for the Starobinsky potential, the termination time of this stage is the same regardless of the $\theta$ value. The behavior of both the scalar field and the evolution of e-folds during this stage mirrors that observed in the super-inflation period.

During the slow-roll inflation phase, an increase in $\theta$ leads to an earlier onset of this stage for the Starobinsky potential. Conversely, for the quadratic potential model, an increase in $\theta$ corresponds to a later commencement of this phase. Analogous to the two preceding stages, the e-folds and the scalar field exhibit consistent behaviors.

Finally, another notable distinction lies in the inflationary exit mechanism. The $\phi^2$ potential model demonstrates an elegant exit from the inflationary stage, whereas such an exit is absent in the Starobinsky potential, leading to eternal inflation. 

\subsection{Starobinsky potential vs. Quadratic potential: KED Regime}

For the commutative case, the super-inflation period ends at very similar times for both models. The maximum value reached by the Hubble parameter is the same in both models, and this similarity also holds for the number of e-folds during this stage. However, for the noncommutative cases, the differences become more pronounced. For the $\phi^2$ potential, the times at which the super-inflation period ends are significantly shorter than those for the Starobinsky potential. The amplitude of the Hubble parameter, and thus its maximum value, increases considerably compared to the Starobinsky potential. This pronounced growth in amplitude is due to the presence of noncommutativity.

During the transition stage, the main difference between the two models is their duration. For the quadratic potential is very short compared to that of the Starobinsky potential, in both the commutative and noncommutative cases, with the difference being of the order $\sim10^6$.

\begin{figure}[H]
\centering
  \includegraphics[scale=1]{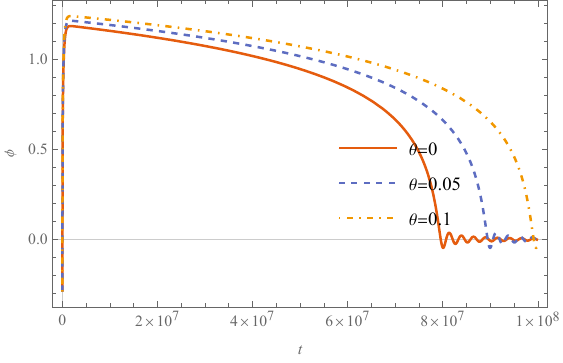}
  \caption{In the figure show the numerical solution of $\phi(t)$ with the initial conditions $\phi_B=-2.8~m_{pl}$ and $\dot\phi_B=0.8~m_{pl}^2$.}
  \label{fig:Newphi_KED}
\end{figure}

For the slow-roll stage we present several precise observations regarding the dynamics. Firstly, an increase in the noncommutative parameter $\theta$ induces a distinct temporal shift in the onset of the slow-roll phase depending on the potential model. Specifically, for the quadratic potential, the slow-roll stage commences earlier as $\theta$ increases. Conversely, the Starobinsky potential exhibits the opposite behavior, with the slow-roll stage shifting to later times as $\theta$ is escalated. Secondly, the number of e-folds demonstrates a sharply contrasting growth rate between the two potentials. The $\phi^2$ potential exhibits a very rapid growth in $N(t)$, whereas the Starobinsky potential shows the opposite, more subdued behavior. Crucially, for both models, the total number of e-folds, is observed to increase with increasing $\theta$. Finally, the mechanisms governing the termination of the slow-roll stage differ significantly. For the $\phi^2$ potential, the slow-roll phase is not observed to end for the commutative case and specific noncommutative values (e.g., $\theta=0.1$), indicating the occurrence of eternal inflation. In contrast, the Starobinsky potential demonstrates a more complex initial evolution for certain cases. For the commutative case $\theta=0$ and $\theta=0.5$, the inflaton field $\phi$ is initiated on the left side of the potential minimum with the specific initial conditions ($\phi=-2.8~m_{pl}$, $\dot\phi=0.8~m_{pl}^2$). Due to these initial conditions, the field first rolls down the potential. Subsequently, the inflaton field crosses the minimum and begins to increase in value, consequently escalating the potential. This initial period of ascent and subsequent decline precedes the actual commencement of the slow-roll inflation (as detailed in Table 2). Following this initial transient phase, the inflaton potential declines monotonically until the end of the inflationary period (as depicted in Fig~\ref{fig:Newphi_KED}).       

\subsection{Starobinsky potential vs. Quadratic potential:  PED Regime}
For this energy regime, the most crucial difference lies in the capability of the respective models to sustained a successful inflationary period consistent with observations. In the chaotic potential model, is demonstrated that an inflationary period is viable, fulfilling the necessary condition of $N(t)\geq 60$. Furthermore, the total number of e-folds, $N(t)$, is found to increase with the value of the noncommutative parameter $\theta$. In sharp contrast, the analysis of the Starobinsky potential model, is consistent with prior research establishing that when the potential energy dominates at the bounce, the universe is unable to transition into a phenomenologically viable inflationary stage.

The introduction of the noncommutative parameter, $\theta$, consistently influences the dynamics in both cases, primarily enhancing the super-inflation phase. In the quadratic potential model, increasing $\theta$ increases the duration of the Super-Inflation (SI) period and the maximum value of the Hubble parameter. Similarly, in the Starobinsky model, the introduction of noncommutative corrections induces an increase in the scale factor.

For the quadratic potential, the duration of the post-bounce stages in the PED regime is significantly less affected by noncommutativity compared to other energy regimes (EKED, KED), with fluctuations in the order of $\sim 10^{-7}$.
\section{Conclusions}\label{conclusions}
In this work, we have examined the (pre)-inflationary dynamics of a flat, homogeneous and isotropic universe within the framework of noncommutative effective loop quantum cosmology, employing the Starobinsky potential as the inflaton model. Through detailed numerical analysis of the modified field equations for various initial conditions and values of the noncommutative parameter, we explored the background evolution across the extreme kinetic-energy domination, kinetic-energy domination, and potential-energy domination regimes, and compared our results with those obtained previously for the quadratic (chaotic) potential in the same formalism \cite{Diaz-Barron:2023ctp}.

Our results demonstrate that the introduction of noncommutativity includes nontrivial modifications in the dynamical evolution near the quantum bounce, affecting both the duration and structure of the inflationary phases. In particular, the Starobinsky potential exhibits an inverse response to noncommutativity compared with the quadratic potential, while noncommutativity tends to prolong the cosmological phases in the latter, it shortens them in the former. Moreover, the amplitude of the Hubble parameter and the evolution of the scalar field decreased with increasing noncommutative parameter, leading to a suppression of the inflationary dynamics relative to the commutative case.

The analysis further reveals that as the noncommutative parameter $\theta$ increases, the onset of the slow-roll phase for the quadratic potential model shifts to earlier times. Conversely, the Starobinsky potential model exhibits the opposite behavior, where the slow-roll phase is initiated at later times as $\theta$ increases. Also, we observe a significantly more rapid growth in the number of e-folds for the $\phi^2$ potential compared to the Starobinsky potential, which demonstrates the inverse behavior. Nevertheless, in both potential models, $N(t)$ exhibits an increase as the noncommutative parameter $\theta$ is augmented. Finally, the $\phi^2$ potential model does not exhibit an end to the slow-roll phase within the observed regime, indicative of eternal inflation (a similar behavior is noted for the case of $\theta=0.1$). In contrast, for the Starobinsky potential, under both the commutative case ($\theta=0$) and the specific noncommutative case of $\theta=0.5$, the inflaton field is observed to commence its evolution from the left side of the potential's minimum. Specifically, using the initial conditions ($\phi_B=-2.8~m_{pl}$, $\dot\phi=0.8~m_{pl}^2$), the inflaton initially rolls down the potential. Following the crossing of the minimum, the inflation's value begins to increase, thereby escalating the potential until the initiation of slow-roll inflation (as detailed in Table \ref{tab:KED_Starobinsky}). Subsequently, the potential decreases as the inflaton rolls until the termination of the inflationary period (as depicted in Figure 12). The potential continues to decline after this point until the end of the inflationary period.

For the PED regime we found that the growth rate of the scale factor is observed to be slower compared to the EKED and KED regimes. A notable finding is that the introduction of noncommutative corrections induces an increase in the scale factor. Another conclusion, consistent with prior research, establishes that when the potential energy dominates at the bounce, the universe is unable to transition into a phenomenologically viable inflationary stage. Detailed numerical simulations, see Table~\ref{tab:PED_Starobinsky}, track key cosmological events. For the commutative case, super-inflation ends relatively early ($t=0.485$ and $N\approx 0.14$ e-folds) , and while the onset of slow-roll inflation is detected at $t=1.134\times 10^3$, the slow-roll phase is not observed to terminate, which reinforces the determination of its non-viability.

From a dynamical systems perspective, the phase space trajectories of the inflaton field reproduce the expected qualitative behavior of LQC models --featuring a nonsingular bounce followed by super-inflation and slow-roll phases-- while preserving consistency with earlier results. Key findings are the identification and stability analysis of the system's fixed points, the first is at $(0, 0)$. Linear stability analysis identified this point as a linear center because its eigenvalues are purely imaginary $\pm i2^{1/4}m$. The second fixed point is at  $(\frac{1}{2}m, 0)$, where the associated eigenvalues are $\{0, \frac{3}{4}m\sqrt{4 - m^2}\}$. With a more comprehensive nonlinear stability analysis, utilizing a Lyapunov function, is concluded that this fixed point is unstable. 

Future research may extend the present analysis by incorporating perturbative degrees of freedom, examining the imprint of noncommutativity on primordial spectra, and exploring the interplay between quantum-geometric corrections and noncommutative deformations in more general inflationary potentials.

\begin{acknowledgements}
S.P.P. was partially funded by SNI-CONACyT and Secretaria de Investigación y Posgrado del Instituto Politécnico Nacional grant SIP20254268. L.R.D.B. was partially funded by SNII-SECIHTI and Secretaría de Investigación y Posgrado del Instituto Politécnico Nacional grant SIP20251224. A.E.G. was partially funded by SNII-SECIHTI and Secretaría de Investigación y Posgrado del Instituto Politécnico Nacional grant 20254597. J.S. was partially funded by SNII-SECIHTI and PRODEP grant UGTO-CA-3
\end{acknowledgements}

\end{document}